\documentclass[prd,floatfix,preprint,tightenlines,amsmath,amssymb,showpacs,superscriptaddress,nofootinbib,twocolumn]{revtex4-1}
\usepackage{amsmath}
\usepackage{graphicx}
\usepackage{epstopdf}
\usepackage{dcolumn}
\usepackage{color}
\usepackage{bm}
\usepackage{overpic}
\usepackage{rotating}
\usepackage{times}
\usepackage{ulem}
\usepackage{multirow}
\usepackage{float}
\usepackage{mathrsfs}
\usepackage{booktabs}
\usepackage{array,booktabs}
\usepackage[mathlines]{lineno} 
\usepackage{lineno} 
\usepackage{epstopdf}
\usepackage{hyperref}
\newcommand{\ee}{e^+e^-}
\newcommand{\jpsi}{J/\psi}
\newcommand{\psip}{\psi(3686)}
\newcommand{\go}{\rightarrow}
\newcommand{\lps}{\bar{\Lambda}\pi\Sigma}
\newcommand{\lpsm}{\bar{\Lambda}\pi^{+}\Sigma^{-}}
\newcommand{\lpsap}{\Lambda\pi^{-}\bar{\Sigma}^{+}}
\newcommand{\lpsp}{\bar{\Lambda}\pi^{-}\Sigma^{+}}
\newcommand{\lpsam}{\Lambda\pi^{+}\bar{\Sigma}^{-}}
\newcommand{\lpsall}{\bar{\Lambda}\pi^{\pm}\Sigma^{\mp} + c.c.}
\newcommand{\cc}{+c.c.}
\newcommand{\ccb}{c\bar{c}}

\begin{document}
	
	\lefthyphenmin=2
	\righthyphenmin=2
	\tolerance=1000
	\uchyph=0
	
	\normalsize
	\parskip=5pt plus 1pt minus 1pt
	
\title{\boldmath Precise measurement of the branching fractions of \texorpdfstring{$\jpsi\go\lpsm\cc$}{} and \texorpdfstring{$\jpsi\go\lpsp\cc$}{}}
	
\author{
	\begin{small}
		\begin{center}
			M.~Ablikim$^{1}$, M.~N.~Achasov$^{5,b}$, P.~Adlarson$^{75}$, X.~C.~Ai$^{81}$, R.~Aliberti$^{36}$, A.~Amoroso$^{74A,74C}$, M.~R.~An$^{40}$, Q.~An$^{71,58}$, Y.~Bai$^{57}$, O.~Bakina$^{37}$
% ...
		\\
		\vspace{0.2cm}
		(BESIII Collaboration)\\
		\vspace{0.2cm} {\it
			$^{1}$ Institute of High Energy Physics, Beijing 100049, People's Republic of China\\
% ...
$^{5}$ Budker Institute of Nuclear Physics SB RAS (BINP), Novosibirsk 630090, Russia\\
% ...
$^{36}$ Johannes Gutenberg University of Mainz, Johann-Joachim-Becher-Weg 45, D-55099 Mainz, Germany\\
			$^{37}$ Joint Institute for Nuclear Research, 141980 Dubna, Moscow region, Russia\\
% ...
$^{40}$ Liaoning Normal University, Dalian 116029, People's Republic of China\\
% ...
$^{57}$ Southeast University, Nanjing 211100, People's Republic of China\\
			$^{58}$ State Key Laboratory of Particle Detection and Electronics, Beijing 100049, Hefei 230026, People's Republic of China\\
% ...
$^{71}$ University of Science and Technology of China, Hefei 230026, People's Republic of China\\
% ...
$^{74}$ University of Turin and INFN, (A)University of Turin, I-10125, Turin, Italy; (B)University of Eastern Piedmont, I-15121, Alessandria, Italy; (C)INFN, I-10125, Turin, Italy\\
			$^{75}$ Uppsala University, Box 516, SE-75120 Uppsala, Sweden\\
% ...
$^{81}$ Zhengzhou University, Zhengzhou 450001, People's Republic of China\\
			\vspace{0.2cm}
% ...
$^{b}$ Also at the Novosibirsk State University, Novosibirsk, 630090, Russia\\
% ...
		}
		\vspace{0.4cm}
			\end{center}
	\end{small}
}

\begin{abstract}

Based on a data sample of $(10087\pm44)\times10^6$ $J/\psi$ events collected with the BESIII detector, the branching fraction of $\jpsi\go\lpsm\cc$ is measured to be $(1.221\pm 0.002\pm 0.038)\times10^{-3}$, and the branching fraction of its isospin partner mode $\jpsi\go\lpsp\cc$ is measured to be $(1.244\pm 0.002\pm 0.045)\times10^{-3}$ with improved precision. Here the first uncertainties are statistical and the second ones systematic. The isospin symmetry of the $\Sigma$ baryon in charmonium hadronic decay and the “$12\%$ rule” are tested, and no violation is found. The potential of using these channels as $\Sigma$ baryon sources for nuclear physics research is studied, and the momentum and angular distributions of these sources are provided.

\end{abstract}

\maketitle

\section{\boldmath Introduction}

    The decay of vector charmonium states, $\jpsi$ and $\psip$, into light hadron final states occurs through the annihilation of $\ccb$ into gluons or virtual photons, providing insights into strong and electromagnetic interactions. However, the nonperturbative nature of these processes makes theoretical calculations challenging. The hadronic decay patterns can be used to guide the theoretical understanding of the decay dynamics for charmonium states.

    Based on the perturbative QCD, the ratios of the $\psip$ to the $\jpsi$ decaying into the same final state ($h$) are almost the same for many different final states, and follow the ``12\% rule"~\cite{Appelquist:1975,Rujula:1975,Brambilla:2010cs}, {\it i.e.}
    
    \begin{equation}
        \begin{split}
        Q_h & = \frac{{\cal B}(\psip\go h)}{{\cal B}(\jpsi\go h)} \\
            & = \frac{{\cal B}(\psip\go \ee)}{{\cal B}(\jpsi\go \ee)} \\
            & = 12\%.
        \end{split}
        \label{eq0}
    \end{equation}
    
    However, some decay modes have been found violating this rule, with a ratio either suppressed or enhanced relative to 12\%. The $\rho\pi$ mode was found to be suppressed by about 2 orders of magnitude~\cite{Franklin:1983, Ablikim:2005, Briere:2005}, whereas the $K_S^0 K_L^0$ mode was enhanced~\cite{BES:2003yxt,CLEO:2006udi,Metreveli:2012tb,BESIII:2017aom} by a factor of 2. Many theoretical models were proposed to solve the ``$\rho\pi$ puzzle" and other related measurements~\cite{PDG:2022}, but none of them is satisfactory~\cite{MoXH:2007}. 

    Among the experimental measurements, the final states with baryons involved follow the ``12\% rule" reasonably well~\cite{PDG:2022}. Whether this phenomenon is universal for all baryonic final states or there are also channels violating the ``12\% rule" yet to be discovered is still an open question. In this article, we extend the study to three-body decays of $\jpsi\to\lpsall$ where two different final states exist corresponding to the isotriplet $\Sigma$ baryons. Our measurements will allow a test of the ``12\% rule" between $\psip$ and $\jpsi$ decays and a test of the isospin symmetry in these two modes. In a previous study, a significant difference in the cross sections of $\ee\to \Sigma^{+}\bar{\Sigma}^{-}$ and $\Sigma^-\bar{\Sigma}^{+}$ has been observed in nonresonant energy region~\cite{Ablikim:2020}. Since there is also an electromagnetic amplitude contributing to $\jpsi$ decays into light hadrons, isospin breaking in $\jpsi\to \bar{\Lambda}\pi\Sigma\cc$ is not unexpected, although the magnitude of the effect is hard to predict.
    
    In addition to the study of the charmonium decay dynamics, a recent proposal suggested to use the tagged $\Sigma$ baryons from $\jpsi\to \bar{\Lambda}\pi\Sigma$ as particle sources to study the $\Sigma$-nucleon interactions with high-precision and supply important data in understanding high density nuclear matter, such as in neutron stars~\cite{Haidenbauer:2013,Vidana:2018,Tolos:2020}. A super $\jpsi$ factory with $\ee$ annihilations at a center-of-mass energy ($\sqrt{s}$) of $3.097$~GeV, which can generate $10^{12}$ or more $\jpsi$ events per year, may provide plenty of (anti)hyperons as sources of baryons to interact with nuclear targets put outside of the beam pipe~\cite{CZYuan:2021,WMSong:2022,Dai:2022}. As the processes of this article function as potential sources of hyperons, it is important to obtain precise measurements of their branching fractions, as well as the momentum and angular distributions of the $\Sigma$ baryons. 
    
    In this article, we report the improved measurement of the processes $\jpsi\go\lpsm\cc$ and $\jpsi\go\lpsp\cc$. The previous results and their average results are summarized in Table~\ref{tab:Old}. Moreover, we test the isospin symmetry in these modes and the ``$12\%$ rule" compared with the branching fractions of $\psip$ decays~\cite{BR:2013}. We also report the momentum and angular distributions of the $\Sigma$ sources generated from these decays. This analysis is carried out by using the large sample of $(10087\pm44)\times10^6$ $\jpsi$ events collected with the BESIII detector~\cite{HXYang:2022}. The data set collected at $\sqrt{s}=3.080~{\rm GeV}$, with an integrated luminosity of 166~${\rm pb^{-1}}$~\cite{HXYang:2022}, is used to estimate the background events coming directly from the $\ee$ annihilation.

    \begin{table*}[htp]
    	\caption{The previous measurement results of branching fractions (${\cal B}$) of $\jpsi\go\lpsm\cc$ and $\jpsi\go\lpsp\cc$. The average results are calculated using uncertainties as weights.}
    	\centering
    	\begin{tabular}{ccc}
    		\hline\hline
			 Reference & ${\cal B}_{\jpsi\go\lpsm\cc}(\times10^{-3})$ & ${\cal B}_{\jpsi\go\lpsp\cc}(\times10^{-3})$ \\ \hline
			MarkII,1984~\cite{Eaton:1983kb} & $1.53\pm0.17\pm0.38$ & $1.38\pm0.21\pm0.35$ \\
			DM2,1987~\cite{DM2:1987nsg} & $0.90\pm0.06\pm0.16$ & $1.11\pm0.06\pm0.20$ \\
			BES,2007~\cite{BR:2007} & - & $1.52\pm0.08\pm0.16$ \\ \hline
			Average & $0.99\pm0.16$ & $1.35\pm0.13$ \\
			\hline\hline
    	\end{tabular}
    	\label{tab:Old}
    \end{table*}

\section{\boldmath BESIII Detector and Monte Carlo Simulation}
\label{sec:data_sets}
	
    The BESIII detector~\cite{Ablikim:2009aa} records symmetric $e^+e^-$ collisions provided by the BEPCII storage ring~\cite{Yu:IPAC2016-TUYA01} in the center-of-mass energy range from 2.0 to 4.95~GeV, with a peak luminosity of $1\times10^{33}\;\text{cm}^{-2}\text{s}^{-1}$ achieved at $\sqrt{s}=3.77\;\text{GeV}$. BESIII has collected large data samples in this energy region~\cite{Ablikim:2019hff}. The cylindrical core of the BESIII detector covers 93\% of the full solid angle and consists of a helium-based multilayer drift chamber~(MDC), a plastic scintillator time-of-flight system (TOF), and a CsI (Tl) electromagnetic calorimeter~(EMC), which are all enclosed in a superconducting solenoidal magnet providing a 1.0~T (0.9~T for 2012 $J/\psi$ data) magnetic field. The solenoid is supported by an octagonal flux-return yoke with resistive plate counter muon identification modules interleaved with steel~\cite{KXHuang:2022}. The charged-particle momentum resolution at $1~{\rm GeV}/c$ is $0.5\%$, and the ${\rm d}E/{\rm d}x$ resolution is $6\%$ for electrons from Bhabha scattering. The EMC measures photon energies with a resolution of $2.5\%$ ($5\%$) at $1$~GeV in the barrel (end cap) region. The time resolution in the TOF barrel region is 68~ps, while that in the end cap region is 110~ps. The end-cap TOF system was upgraded in 2015 using multigap resistive plate chamber technology, providing a time resolution of 60~ps~\cite{XLi:2017,YXGuo:2017,PCao:2020}.
    
    Simulated data samples produced with a {\sc GEANT4}-based~\cite{geant4:2003} Monte Carlo (MC) package, which includes the geometric description of the BESIII detector and the detector response, are used to determine detection efficiencies and to estimate backgrounds. The inclusive MC sample includes both the production of the $\jpsi$ resonance and the continuum processes incorporated in {\sc KKMC}~\cite{kkmc:2000,kkmc:2001}. All particle decays are modeled with {\sc EvtGen}~\cite{evtgen:2001,evtgen:2008} using branching fractions either taken from the Particle Data Group (PDG)~\cite{PDG:2022}, when available, or otherwise estimated with {\sc LundCharm}~\cite{lundcharm:2000,lundcharm:2014}. Final-state radiation from charged final-state particles is incorporated using the {\sc PHOTOS} package~\cite{photos:1993}. For each signal process, an exclusive phase-space MC sample of $4\times10^{5}$ $J/\psi$ events for each mode is generated to optimize the selection criteria. These events are generated according to the phase-space distribution, and weighted later to take into account intermediate states. 

\section{\boldmath Event Selection}
\label{sec:evtsel}
    
    In this analysis, the charge-conjugate reaction is always implied unless explicitly mentioned. The decays $\jpsi\go\lpsm\cc$ and $\jpsi\go\lpsp\cc$ are reconstructed by detecting the decay $\Lambda\go p \pi^-$ and a $\pi^{\pm}$ from $\jpsi$ decays. The $\Sigma$ baryon is not reconstructed from its decay product, but inferred by checking the recoiling mass distribution of the $\bar{\Lambda}\pi$ system, $M_{\rm recoil}(\bar{\Lambda}\pi)$. The following basic selection criteria, including charged track selection, particle identification (PID) and $\Lambda$ reconstruction, are used.
    
    Charged tracks detected in the MDC are required to be within a polar angle ($\theta$) range of $|\rm{cos\theta}|<0.93$, where $\theta$ is defined with respect to the $z$ axis, which is the symmetry axis of the MDC. Candidate events must have three charged tracks with one positively charged track and one negatively charged track at least. PID for charged tracks combines measurements of the d$E$/d$x$ in the MDC and the flight time in the TOF to form likelihoods ($\mathcal{L}$) for charged proton, kaon and pion hypotheses. Tracks are identified as protons when the proton hypothesis has the greatest likelihoods [$\mathcal{L}(p)>\mathcal{L}(K)$ and $\mathcal{L}(p)>\mathcal{L}(\pi)$] and satisfies $\mathcal{L}(p)>0.001$. Charged pions are identified by comparing the likelihoods for the pion hypotheses $\mathcal{L}(\pi)>\mathcal{L}(p)$ and $\mathcal{L}(\pi)>\mathcal{L}(K)$, and required the likelihoods $\mathcal{L}(\pi)>0.001$.
    
\begin{figure*}[htbp]
        \centering
        \includegraphics[width=2.5in]{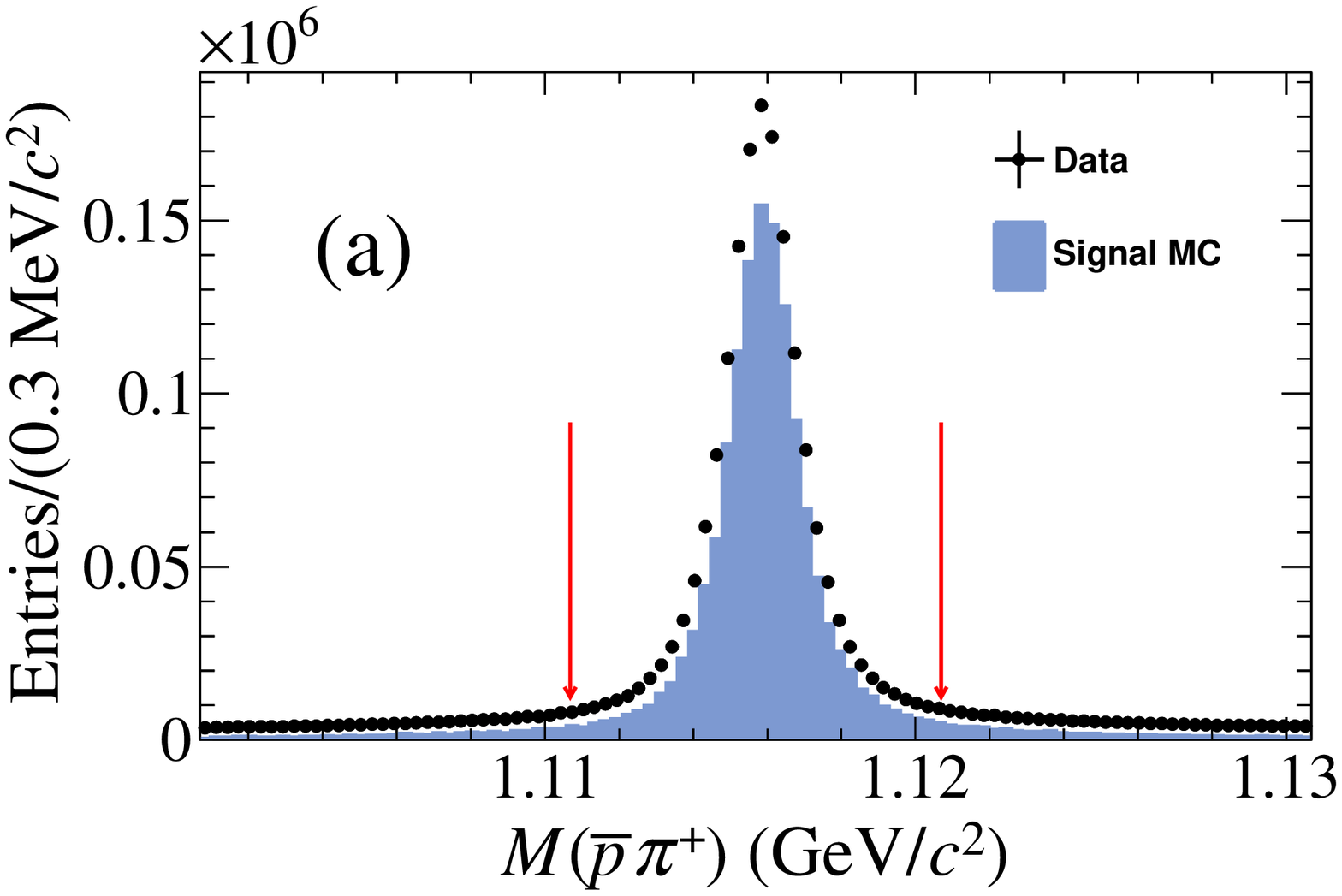}
        \includegraphics[width=2.5in]{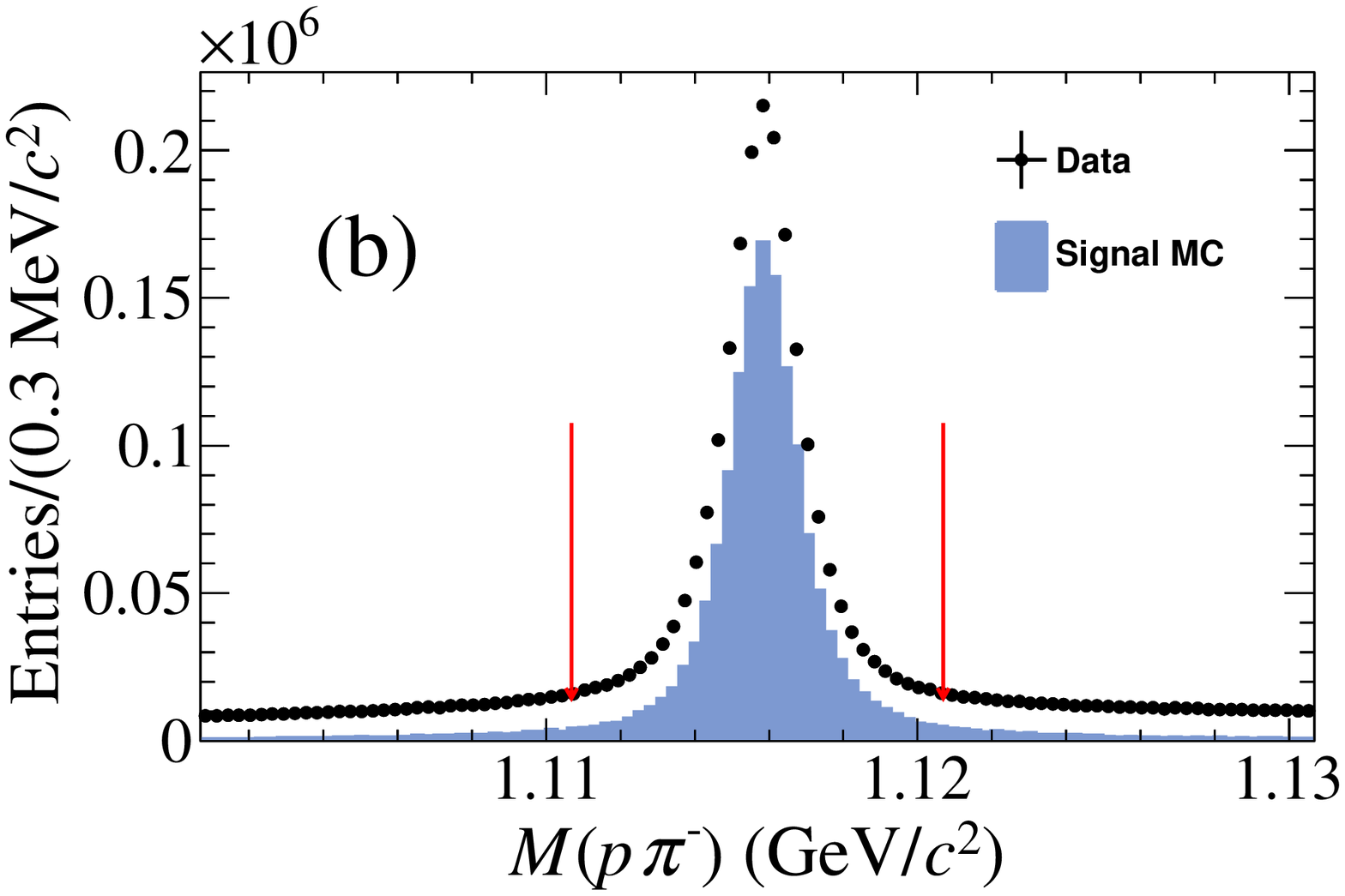}
        \includegraphics[width=2.5in]{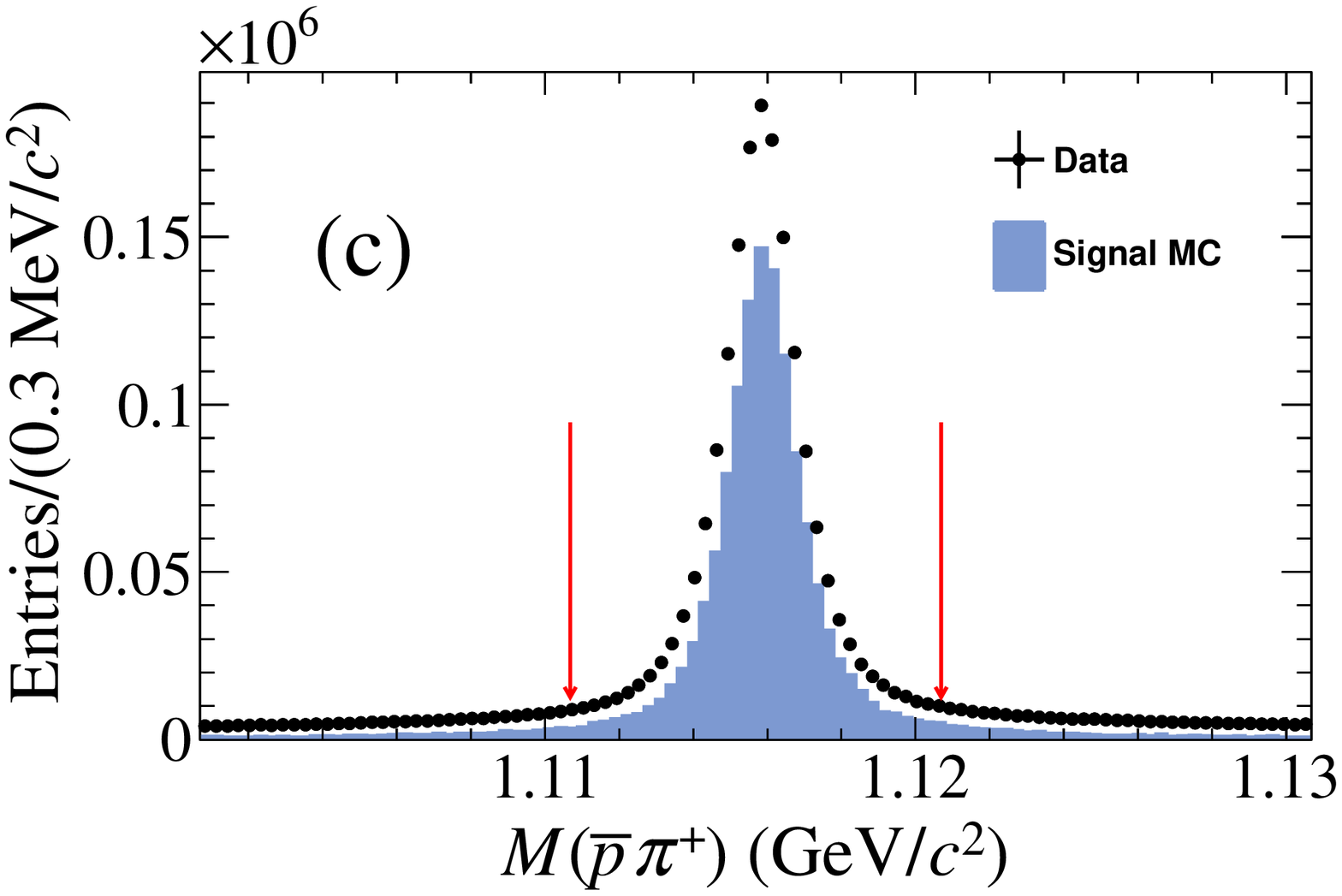}
        \includegraphics[width=2.5in]{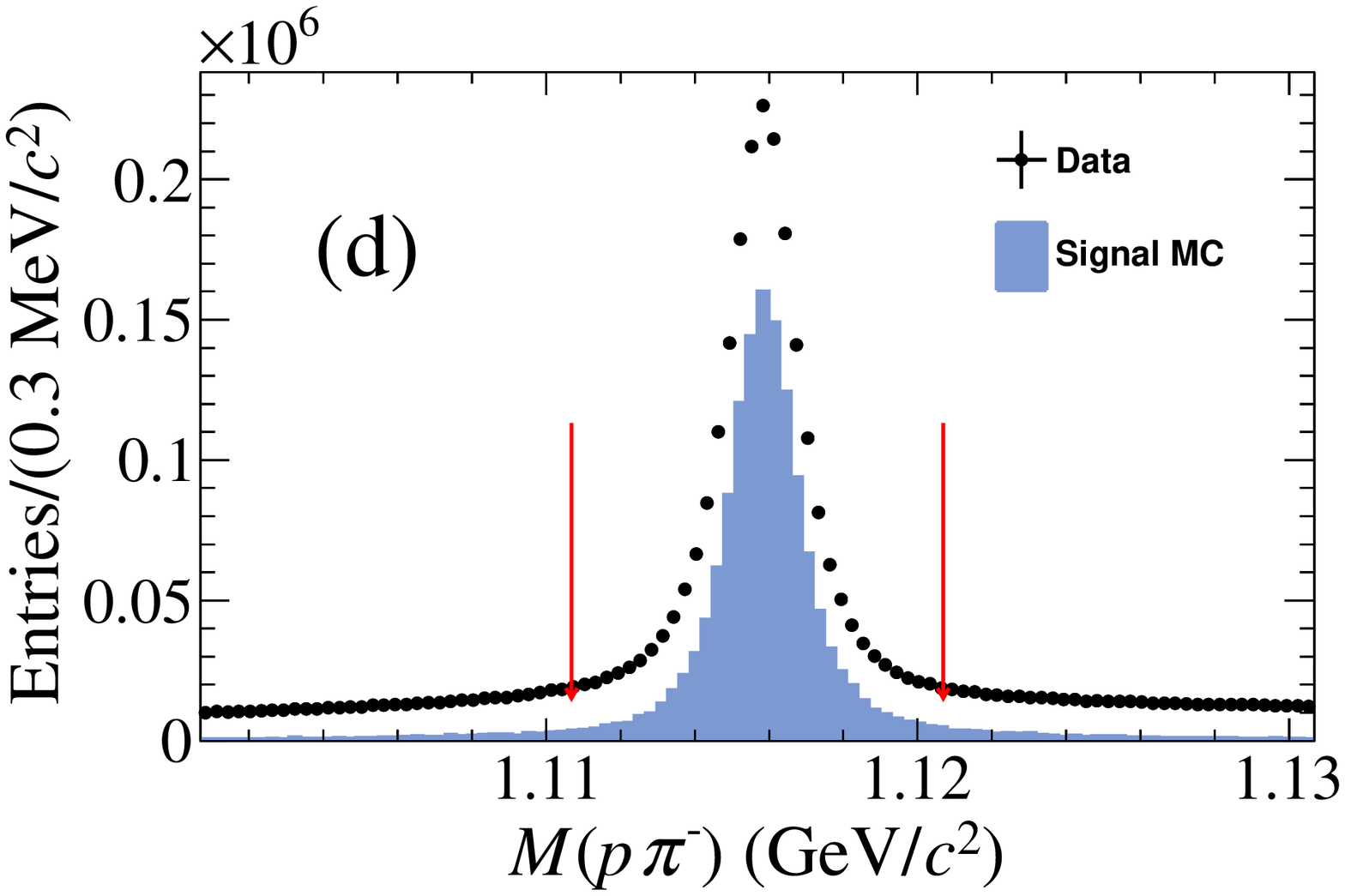}
        \caption{           	
            The $M(p\pi^{-})$ distributions. (a) is for $\bar{\Lambda}$ candidates from $J/\psi\rightarrow\bar{\Lambda}\pi^{+}\Sigma^{-}$. (b) is for $\Lambda$ from $J/\psi\rightarrow\Lambda\pi^{-}\bar{\Sigma}^{+}$. (c) is for $\bar{\Lambda}$ from $J/\psi\rightarrow\bar{\Lambda}\pi^{-}\Sigma^{+}$, and (d) is for $\Lambda$ from $J/\psi\rightarrow\Lambda\pi^{+}\bar{\Sigma}^{-}$. The blue histograms are from signal MC samples, scaled to $N_{\jpsi}\cdot{\cal B}(J/\psi\rightarrow\bar{\Lambda}\pi\Sigma)\cdot{\cal B}(\Lambda\rightarrow p \pi^{-})$, where $N_{\jpsi}$ is the number of $\jpsi$ events in data. The red arrows denote the $\Lambda$ mass window.
        }
        \label{fig:lambda}
    \end{figure*}

    Since $\Lambda$ has relatively long lifetime, it is reconstructed by constraining the $p\pi^{-}$ pair to a secondary vertex. The decay length of $\Lambda$ from the secondary vertex fit divided by its corresponding uncertainty is required to be greater than two. If more than one $\Lambda$ candidate survives, the one with the minimum value of $[M(p\pi^{-})-m(\Lambda)]^2$ is kept for further analysis, where $M(p\pi^{-})$ is the invariant mass of $p\pi^{-}$ and $m(\Lambda)$ is the known mass of $\Lambda$~\cite{PDG:2022}. Figure~\ref{fig:lambda} shows the $M(p\pi^{-})$ distributions, where $\Lambda$ candidates are selected by requiring $|M(p\pi^{-})-m(\Lambda)|<5~{\rm MeV}/c^2$.
    
    For the charged track not originating from $\Lambda$ decays, {\it i.e.,} $\pi^{\pm}$ from $J/\psi$, the distance of closest approach to the interaction point must be less than 10 cm along the $z$ axis, and less than 1 cm in the transverse plane. If there are multiple $\pi^{\pm}$ tracks from $J/\psi$ decays, all of them are saved. The candidates are selected by requiring $M_{\rm recoil}(\bar{\Lambda}\pi)$ to be within the $\Sigma$ mass window [$1.14<M_{\rm recoil}(\bar{\Lambda}\pi)<1.24~{\rm GeV}/c^2$].
    
    By analyzing the $\jpsi$ inclusive MC sample with the topology analysis program TopoAna~\cite{topo:2021}, the background shape within the $\Sigma$ mass window is found to be smooth. Therefore, a third-order Chebyshev polynomial is used to describe the background shape when extracting the number of signal events.
    
    To investigate possible backgrounds from continuum processes $\ee\go\gamma^{*}\go\lps$, the same selection criteria are applied to the data taken at $\sqrt{s}=3.080~{\rm GeV}$. After normalizing to the integrated luminosity of $J/\psi$ data, the number of signal events and the influence on the branching fraction measurements are calculated. Details are shown in Sec.~\ref{sec:BR}.

\section{\boldmath Detection efficiency}
\label{sec:eff}

\begin{figure*}[htbp]
        \centering
        \includegraphics[width=2.5in]{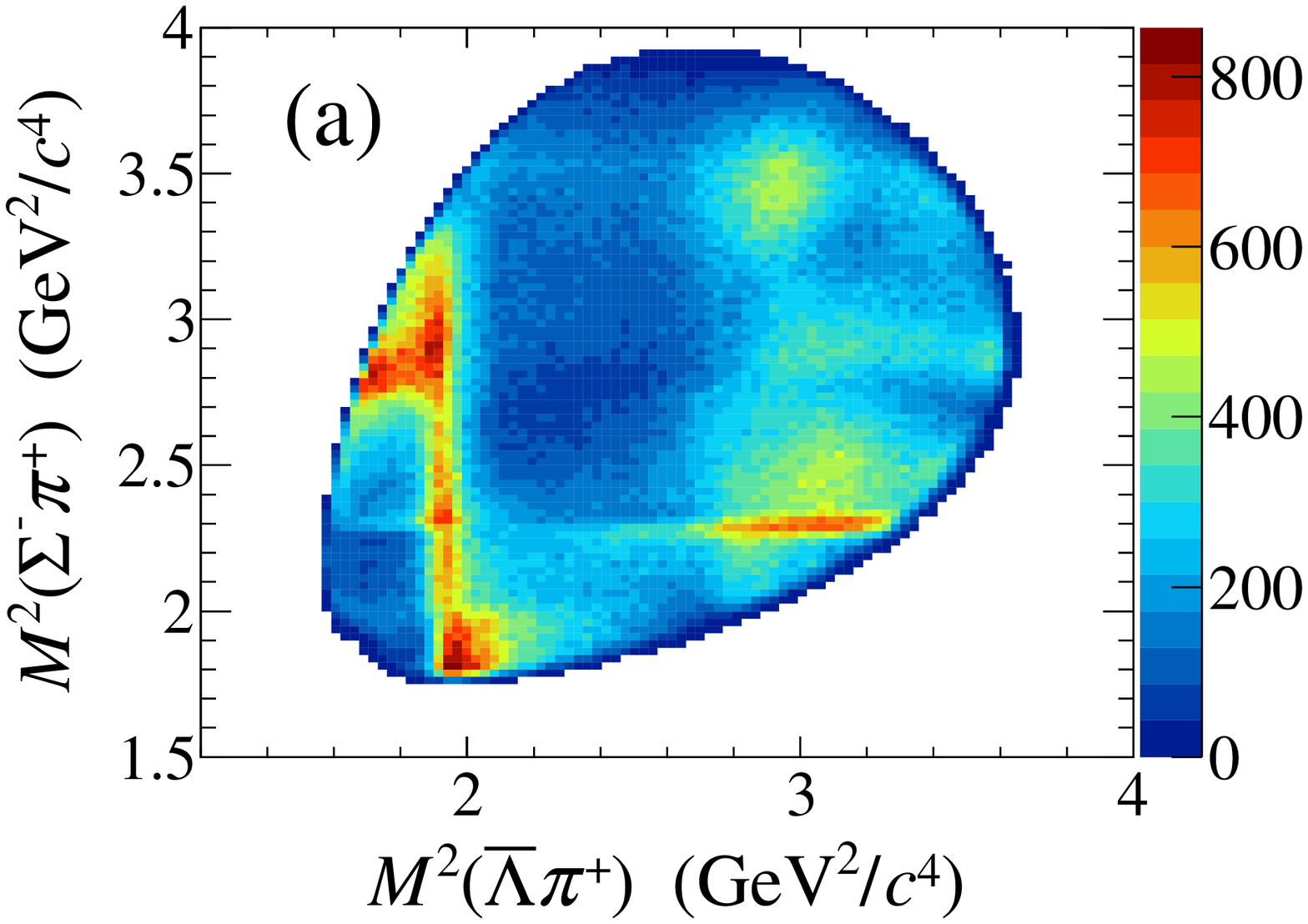}
        \includegraphics[width=2.5in]{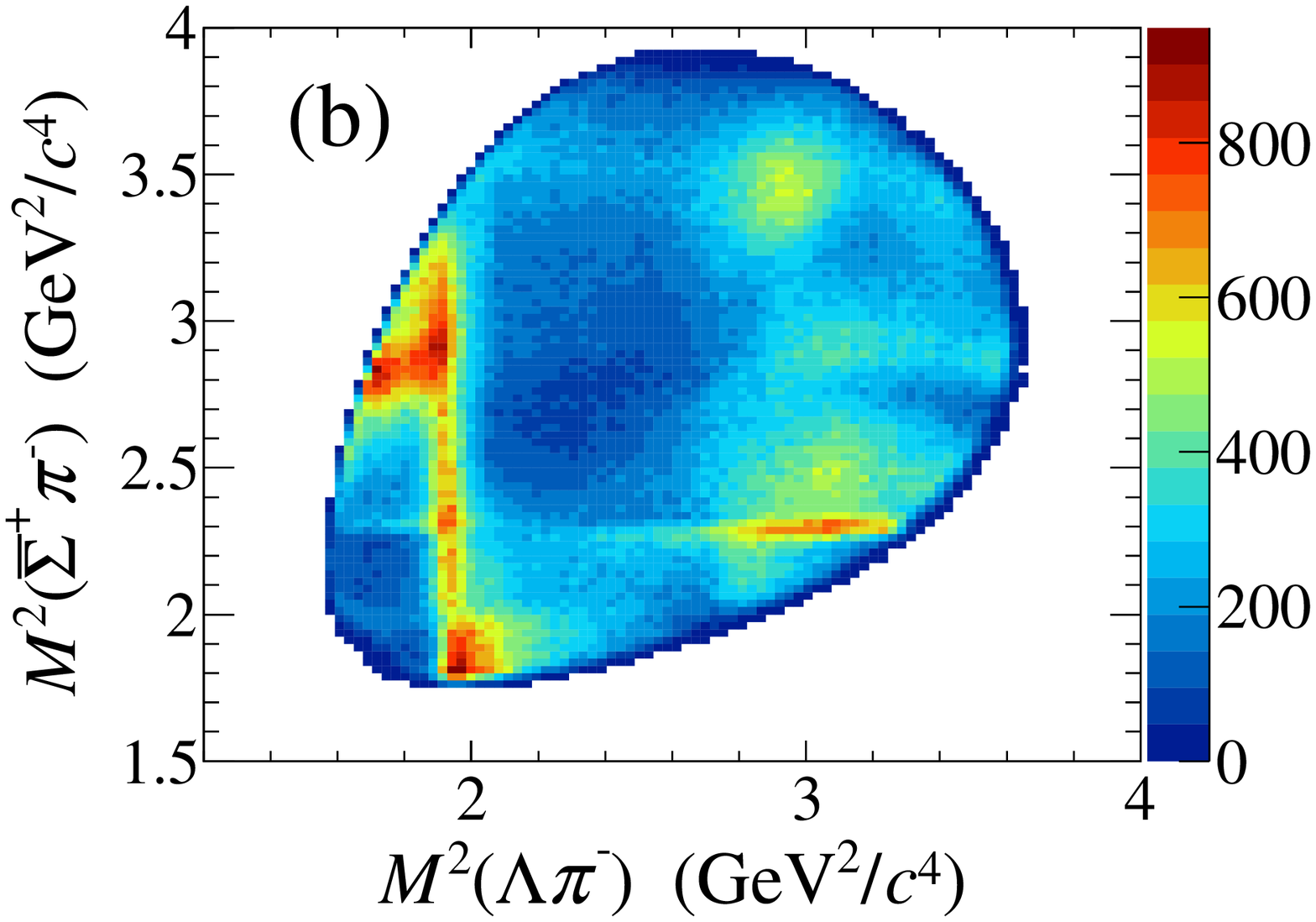}
        \includegraphics[width=2.5in]{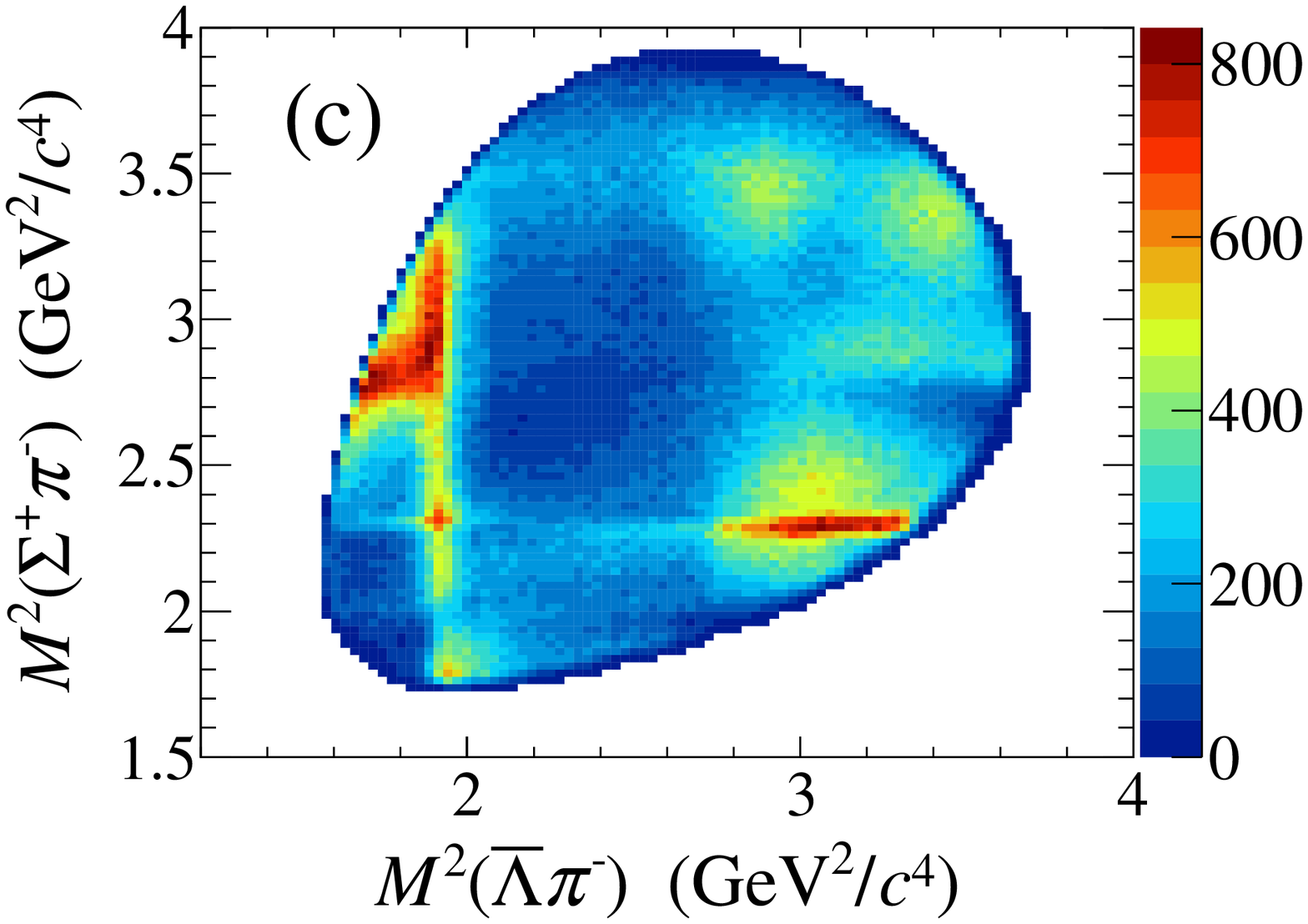}
        \includegraphics[width=2.5in]{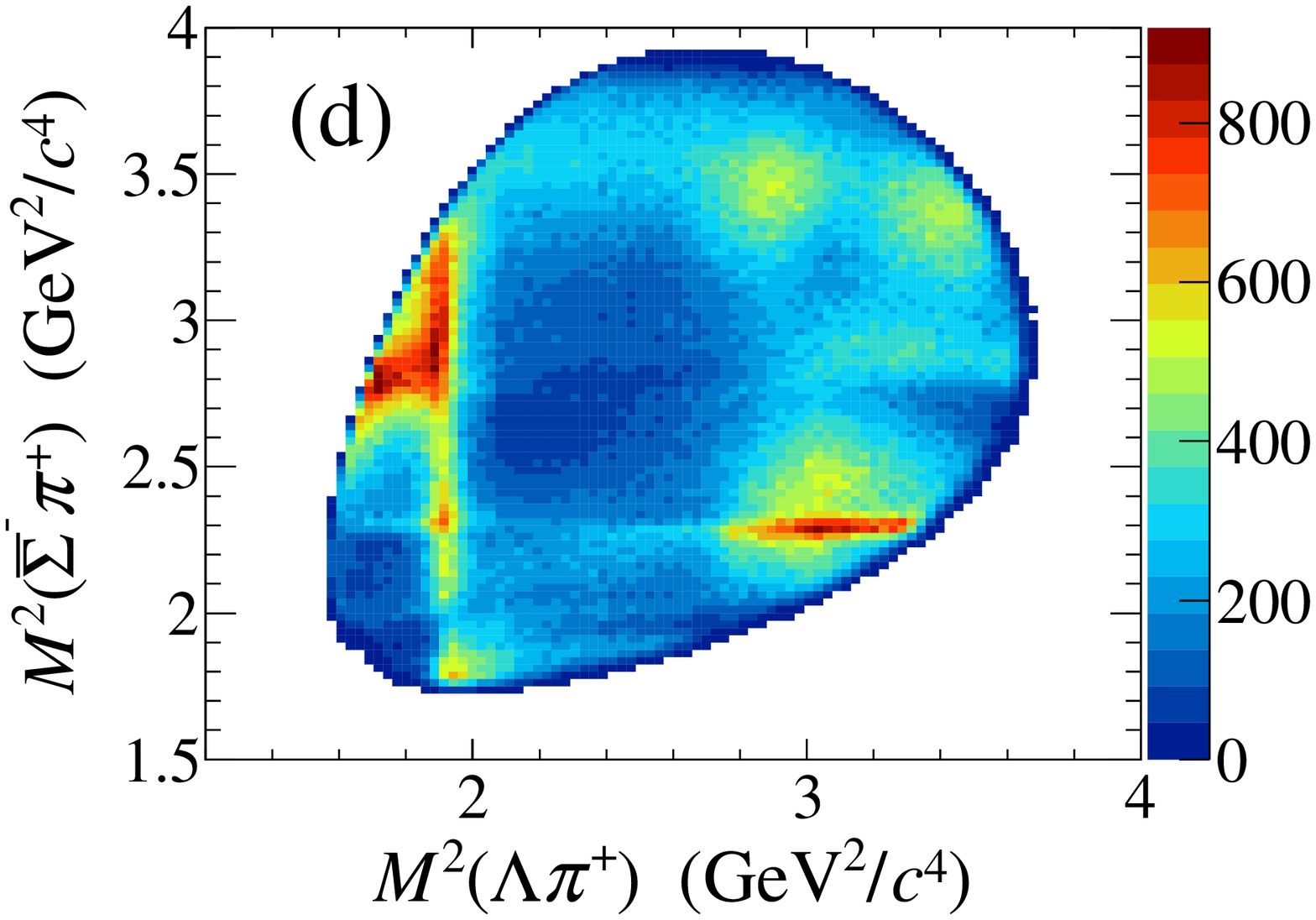}
        \caption{
            The Dalitz plots of $J/\psi\rightarrow\bar{\Lambda}\pi^{+}\Sigma^{-}$ (a), $J/\psi\rightarrow\Lambda\pi^{-}\bar{\Sigma}^{+}$ (b), $J/\psi\rightarrow\bar{\Lambda}\pi^{-}\Sigma^{+}$ (c), and $J/\psi\rightarrow\Lambda\pi^{+}\bar{\Sigma}^{-}$ (d) events in the $\Sigma$ signal region from data.
        }
        \label{fig:dalitz}
    \end{figure*}

    Possible intermediate states decaying into the same final states may affect the detection efficiency. Figure~\ref{fig:dalitz} shows the Dalitz plots of the four modes, where clear $\Lambda^{*}$ and $\Sigma^{*}$ resonances can be seen. To achieve an efficient description of data with intermediate states, a partial-wave analysis (PWA) is applied to the events. To have a purer data sample for the PWA, the signal candidates are selected by reconstructing the full decay chains, which include the decay $\Sigma\go n\pi$. The event selection requirements are similar to those described in Sec.~\ref{sec:evtsel}. For both $\jpsi\go\lpsm$ and $\jpsi\go\lpsp$ decay modes, a one-constraint (1C) kinematic fit is performed by missing a neutron track for the corresponding $\jpsi\go\bar{\Lambda}\pi^{+}\pi^{-}n$ and $\jpsi\go\bar{\Lambda}\pi^{-}\pi^{+}n$ hypotheses, respectively, and similarly for other modes. If there is more than one candidate, the combination with the least $\chi^{2}$ of the 1C kinematic fit ($\chi^{2}_{\rm 1C}$) is selected. In the event-level selection, except the requirements introduced in Sec.~\ref{sec:evtsel}, the $\chi^{2}$ of the vertex fit in the $\Lambda$ reconstruction $\chi^{2}_{\rm vtx}<30$ and kinematic fit $\chi^{2}_{\rm 1C}<30$ are required. To estimate the background contributions, the $\Sigma$ signal and sideband regions for $J/\psi\rightarrow\bar{\Lambda}\pi^{+}\Sigma^{-} + c.c.$ are chosen as $M(n\pi^{-})\in[1.185, 1.205]~{\rm GeV}/c^2$ and $[1.160, 1.170]\cup [1.220, 1.230]~{\rm GeV}/c^2$, respectively. The $\Sigma$ signal and sideband regions for $J/\psi\rightarrow\bar{\Lambda}\pi^{-}\Sigma^{+} + c.c.$ are chosen as $M(n\pi^{+})\in[1.177, 1.197]~{\rm GeV}/c^2$ and $[1.152, 1.162]\cup [1.212, 1.222]~{\rm GeV}/c^2$, respectively. The information of PWA input is shown in Table~\ref{tab:PWA}.
    
    \begin{table*}[htbp]
    	\caption{Information of PWA input: the number of events in the $\Sigma$ signal region of data~($N_{\rm signal}$), the number of background events in data estimated by $\Sigma$ sideband regions~($N_{\rm sideband}$), and the ratios between them.}
    	\centering
    	\small
    	\begin{tabular}{cccc}
    		\hline\hline
    		Decay mode & $N_{\rm signal}$ & $N_{\rm sideband}$ & $N_{\rm sideband}$/$N_{\rm signal}$ \\ \hline
    		$J/\psi\rightarrow\bar{\Lambda}\pi^{+}\Sigma^{-}$ & 130,590 & 3,492 & $2.7\%$ \\
    		$J/\psi\rightarrow\Lambda\pi^{-}\bar{\Sigma}^{+}$ & 139,310 & 4,174 & $3.0\%$ \\
    		$J/\psi\rightarrow\bar{\Lambda}\pi^{-}\Sigma^{+}$ & 70,405  & 4,553 & $6.5\%$ \\
    		$J/\psi\rightarrow\Lambda\pi^{+}\bar{\Sigma}^{-}$ & 75,610  & 4,867 & $6.4\%$ \\ \hline\hline
    	\end{tabular}
    	\label{tab:PWA}
    \end{table*}
    
\begin{figure*}[htbp]
        \centering
        \includegraphics[width=2.0in]{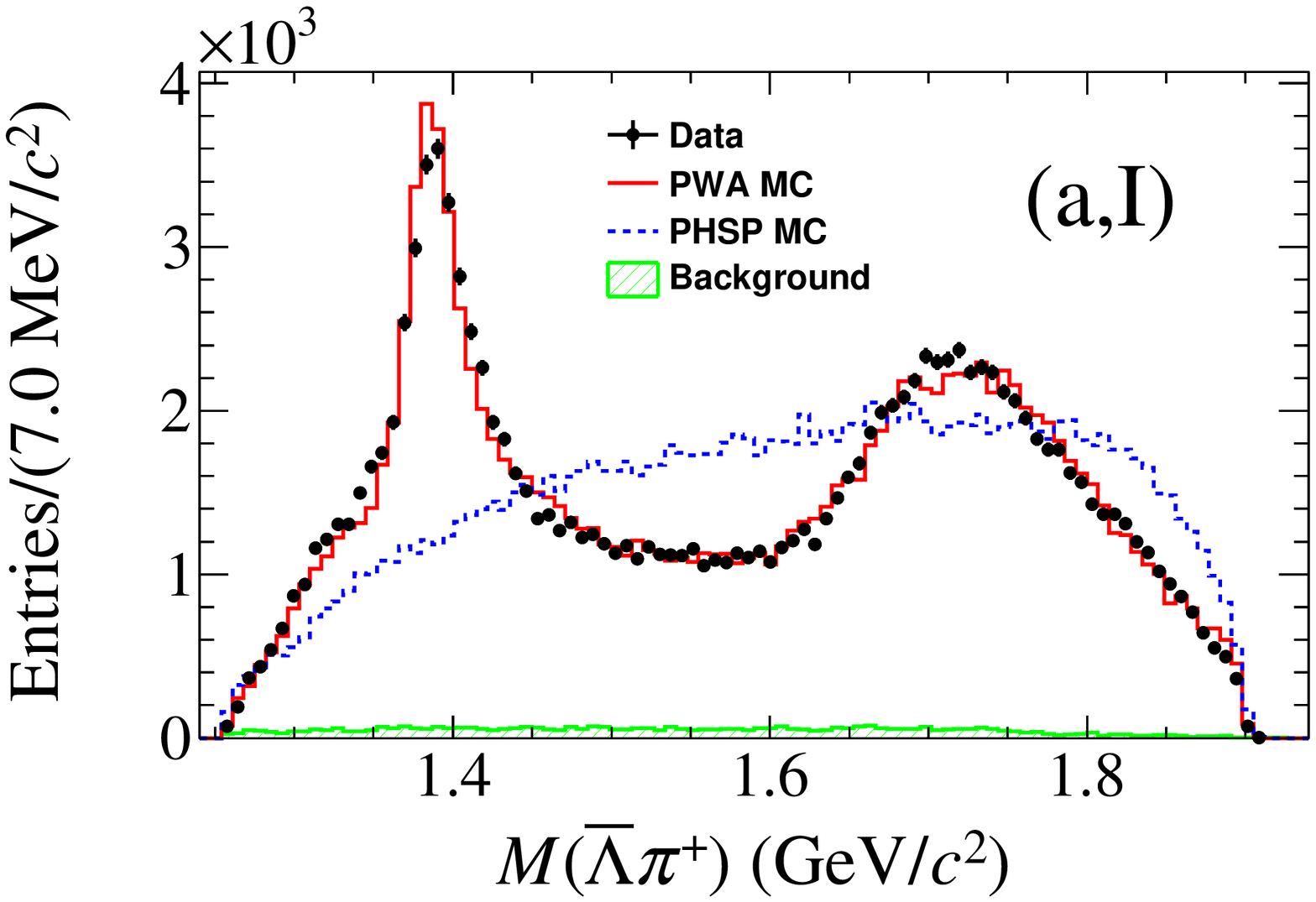}
        \includegraphics[width=2.0in]{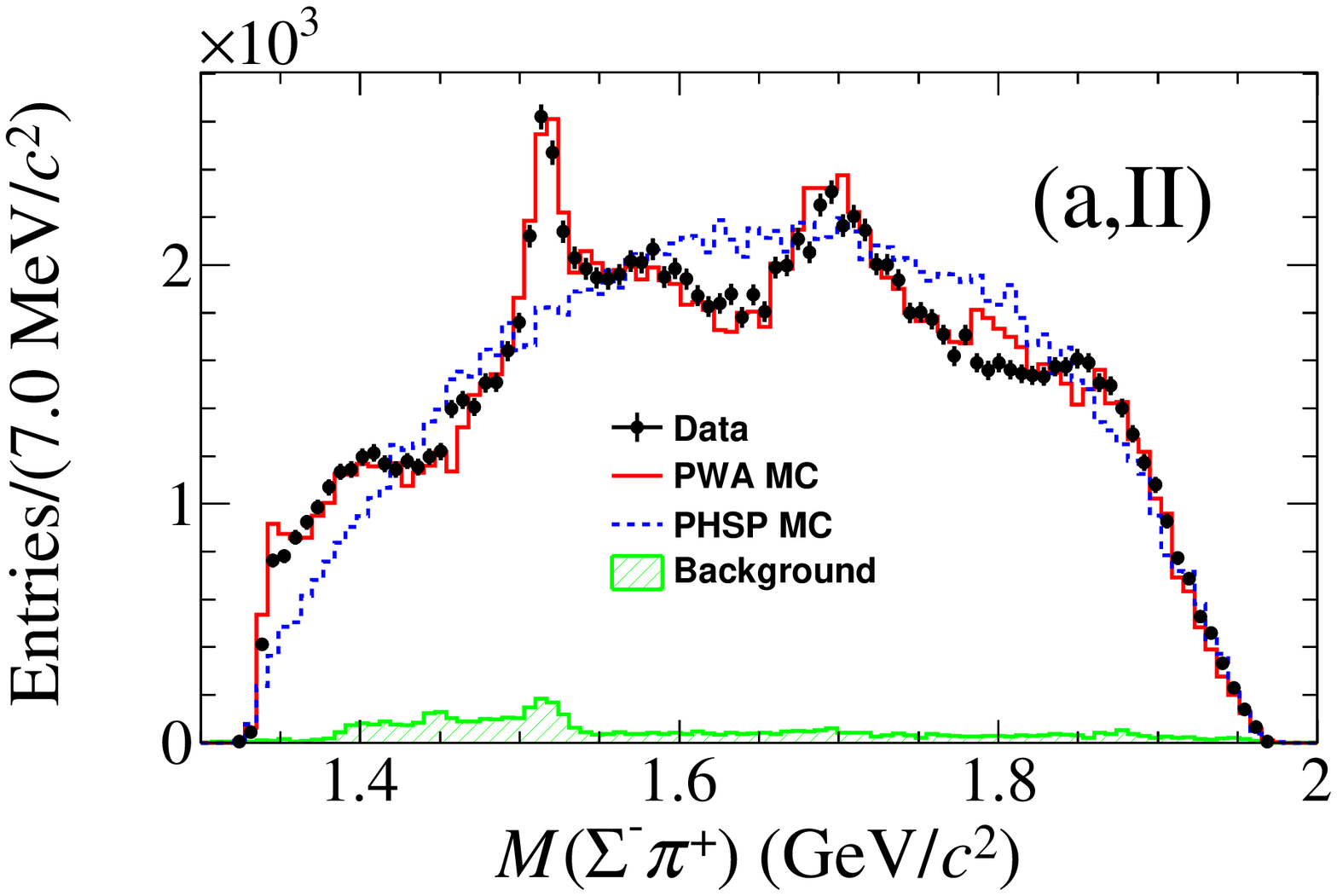}
        \includegraphics[width=2.0in]{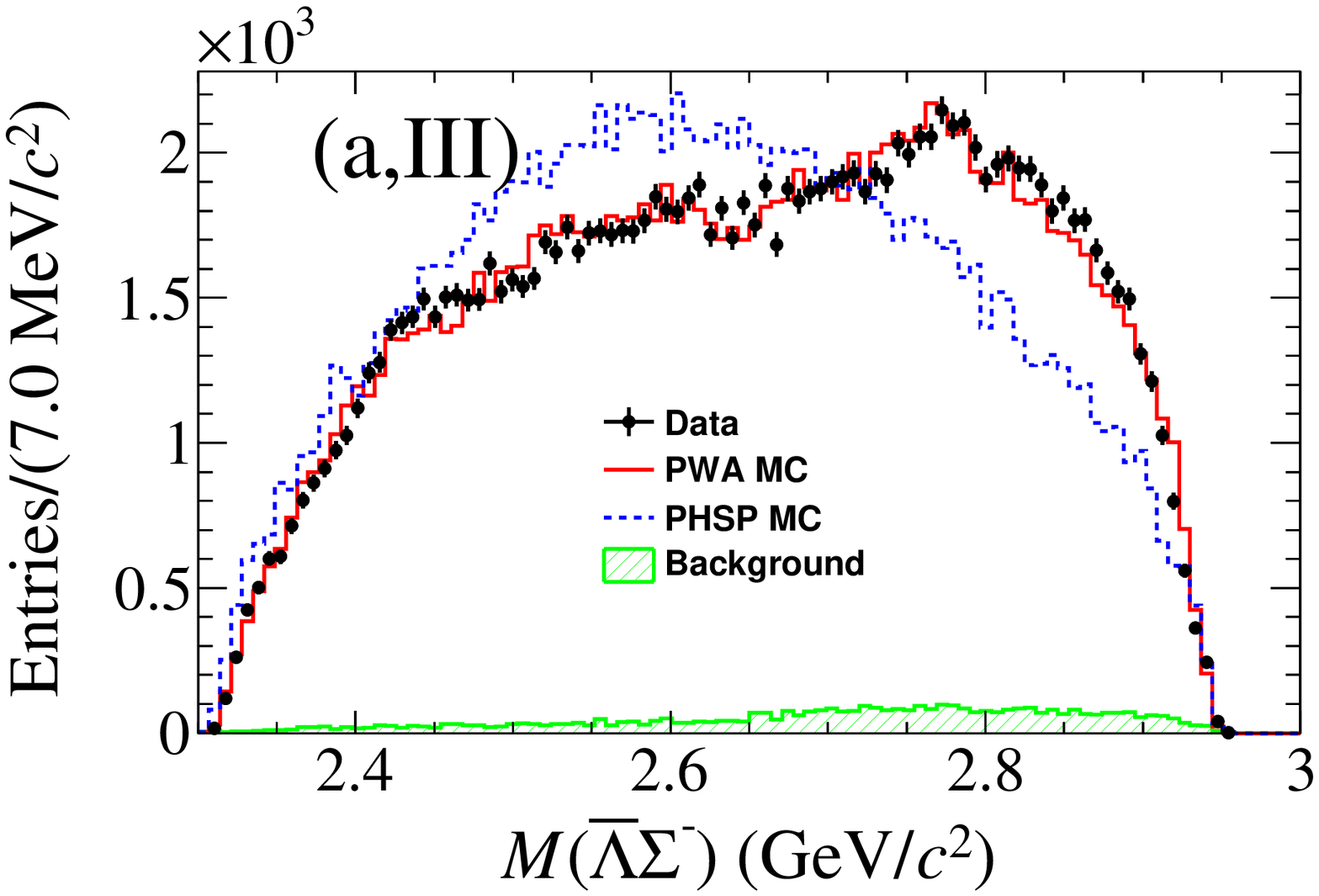}
        
        \includegraphics[width=2.0in]{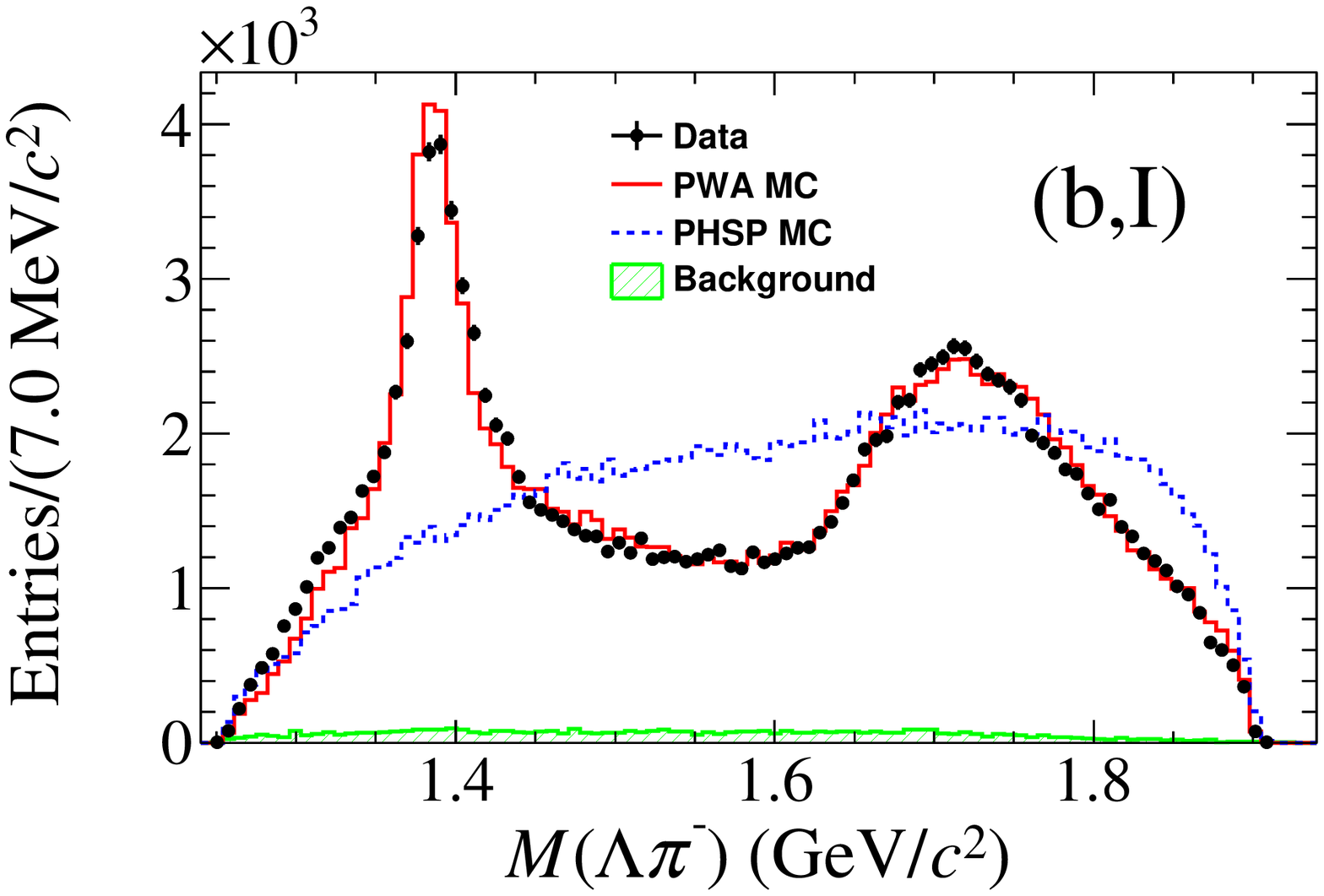}
        \includegraphics[width=2.0in]{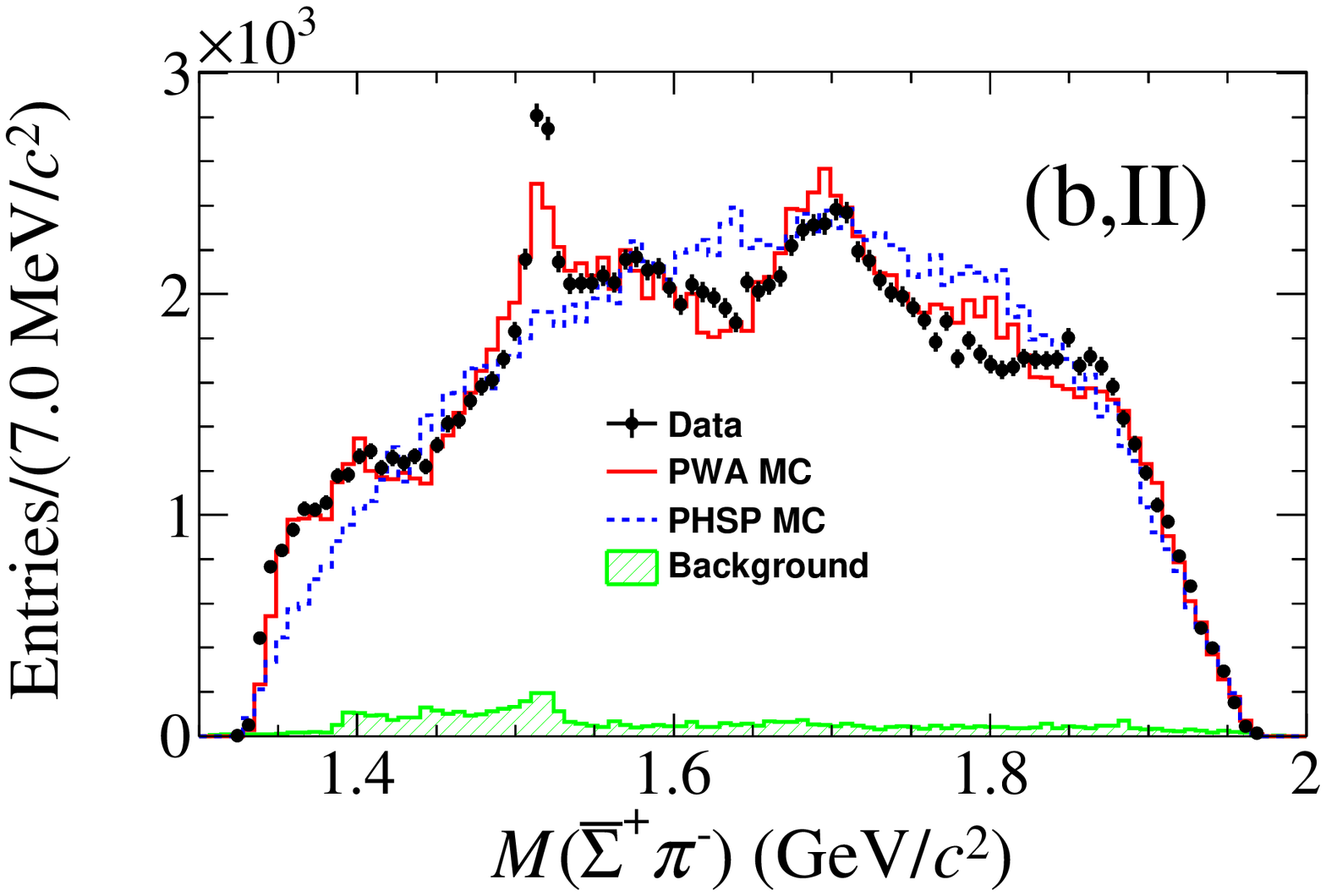}
        \includegraphics[width=2.0in]{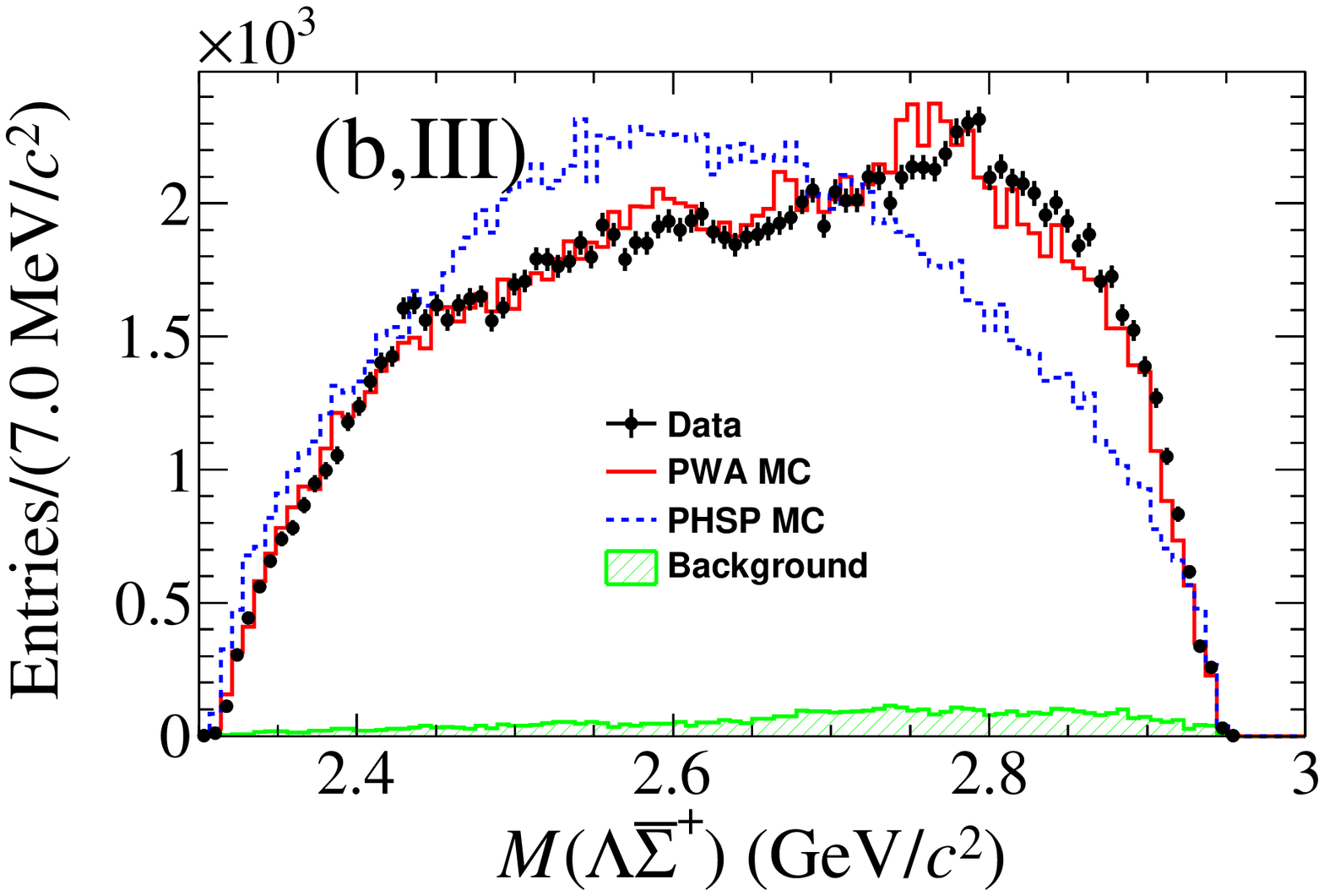}
        
        \includegraphics[width=2.0in]{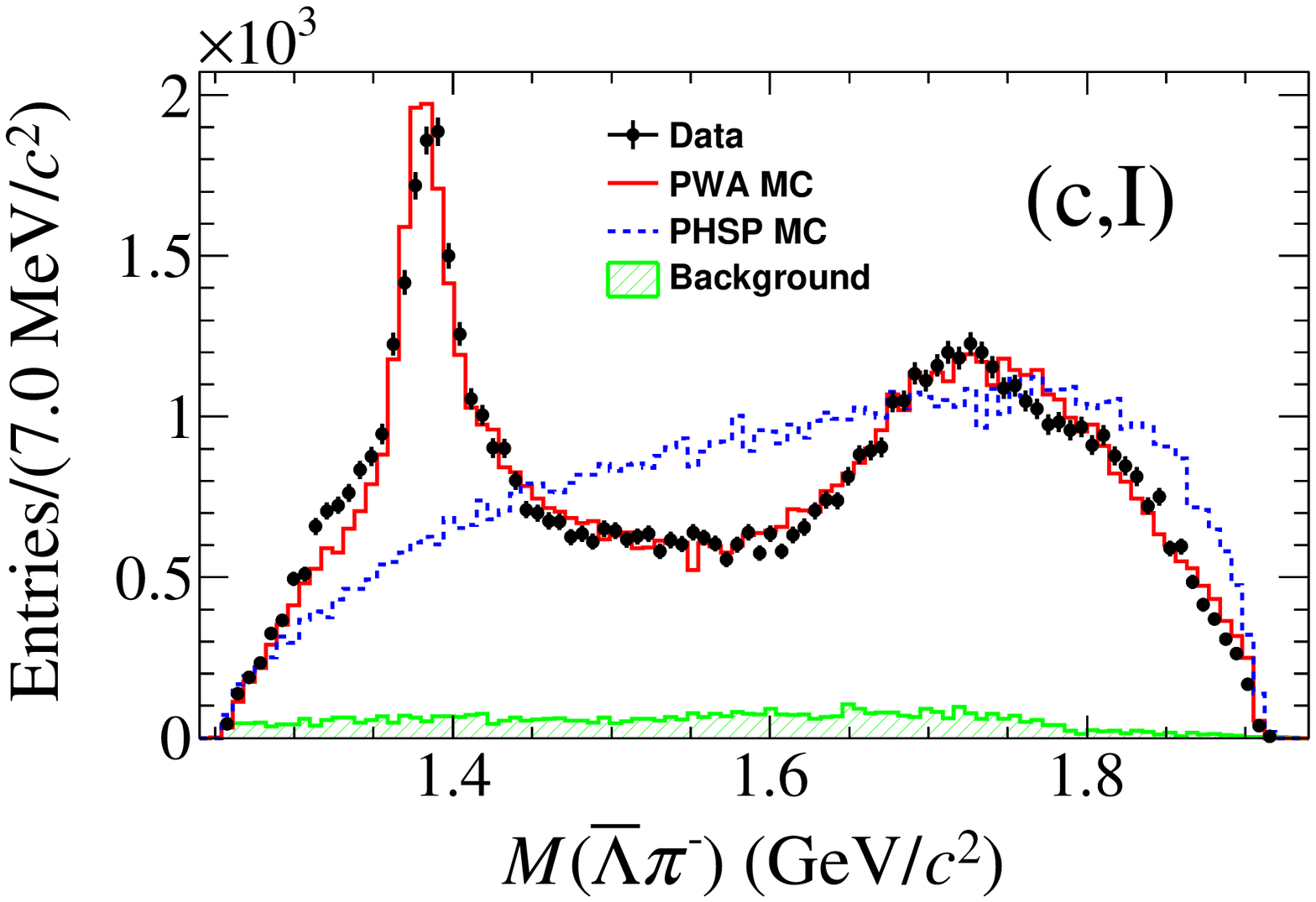}
        \includegraphics[width=2.0in]{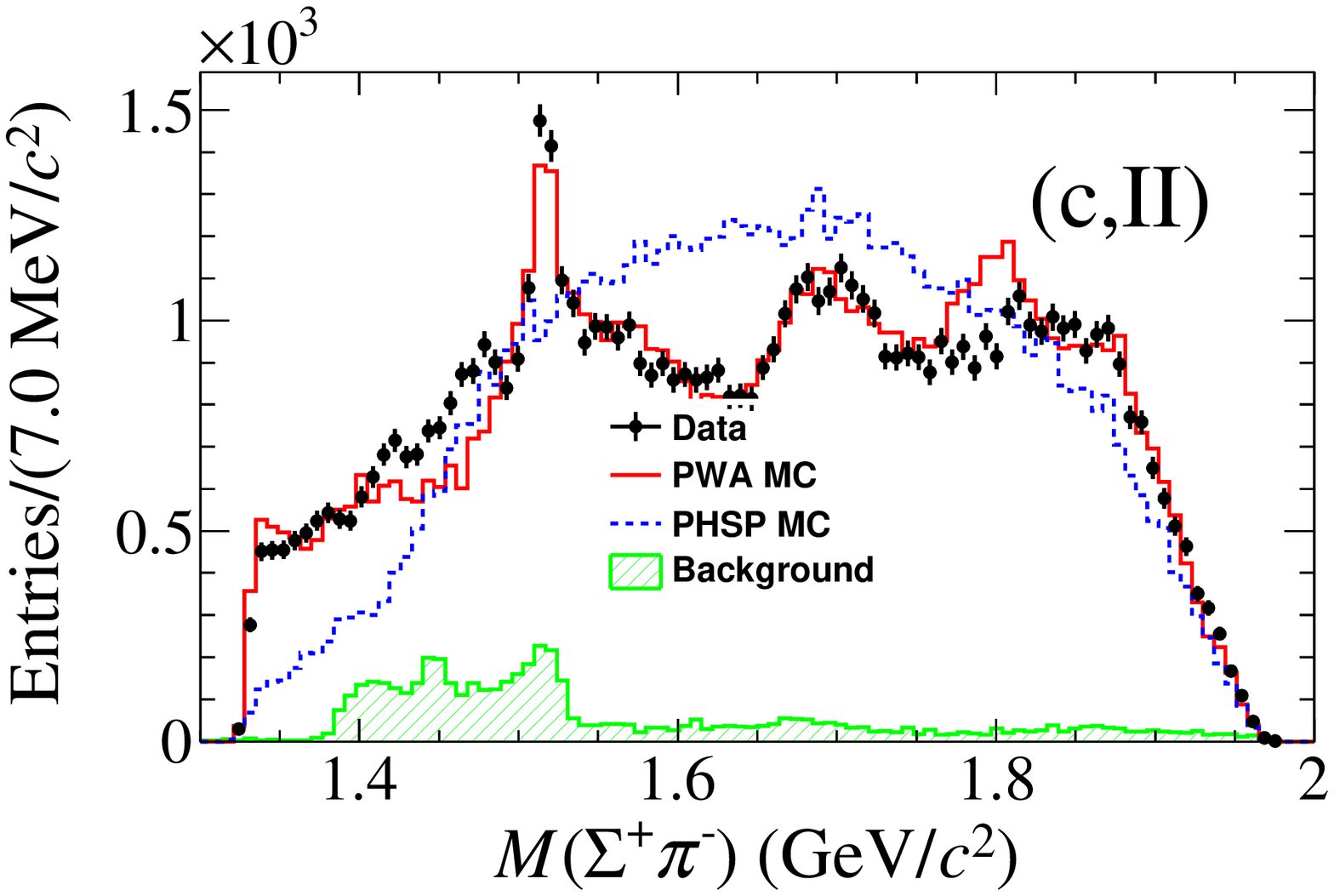}
        \includegraphics[width=2.0in]{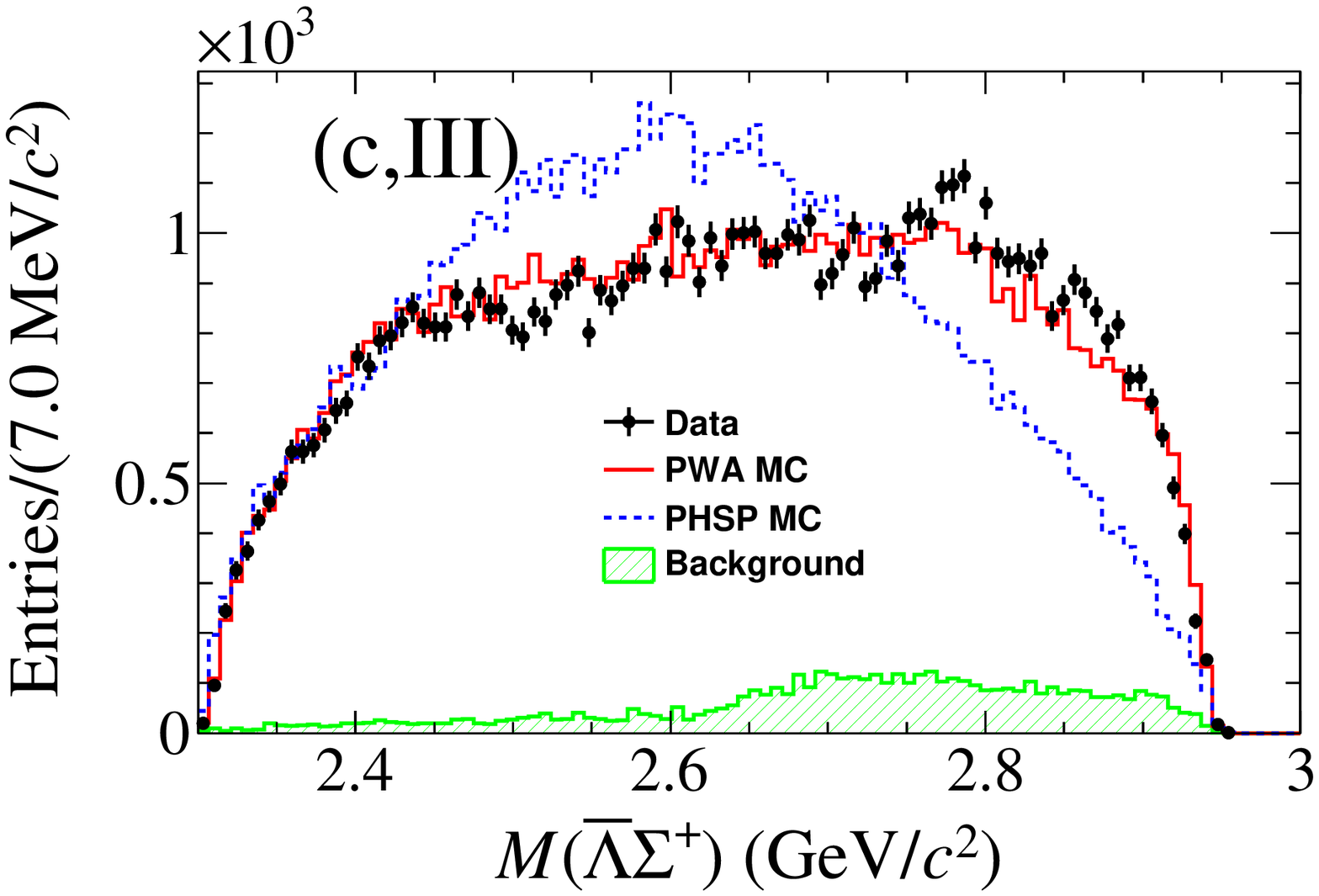}
        
        \includegraphics[width=2.0in]{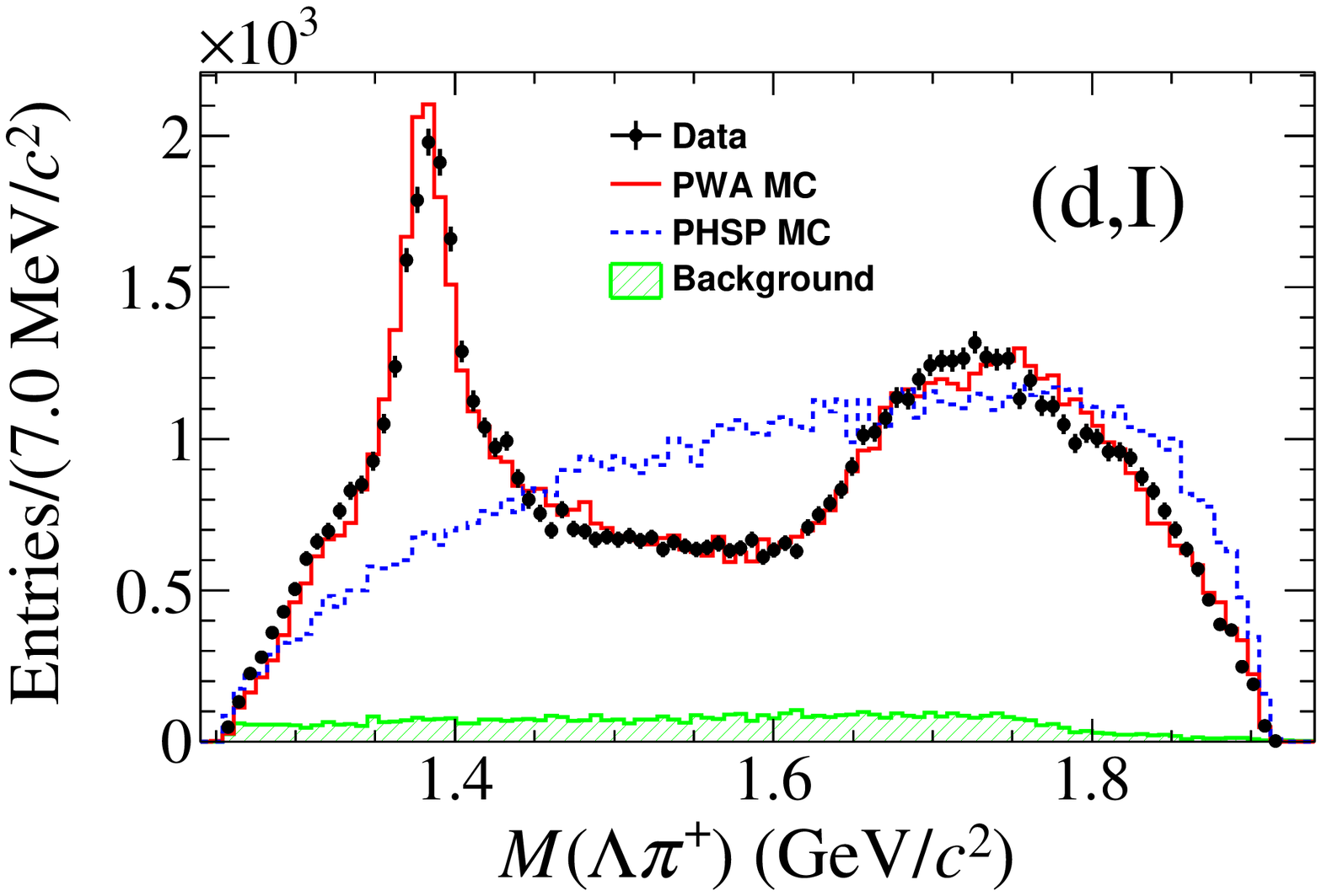}
        \includegraphics[width=2.0in]{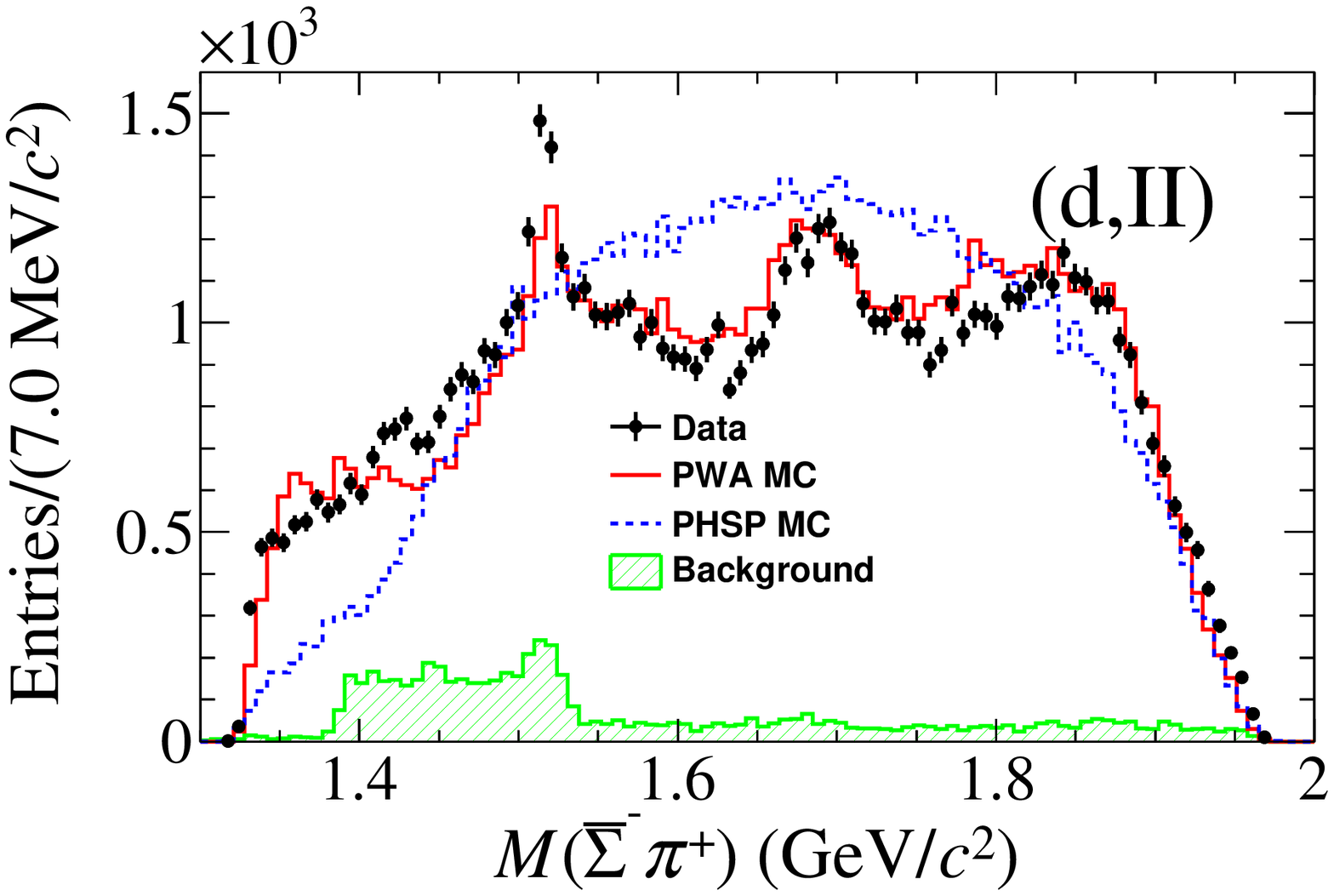}
        \includegraphics[width=2.0in]{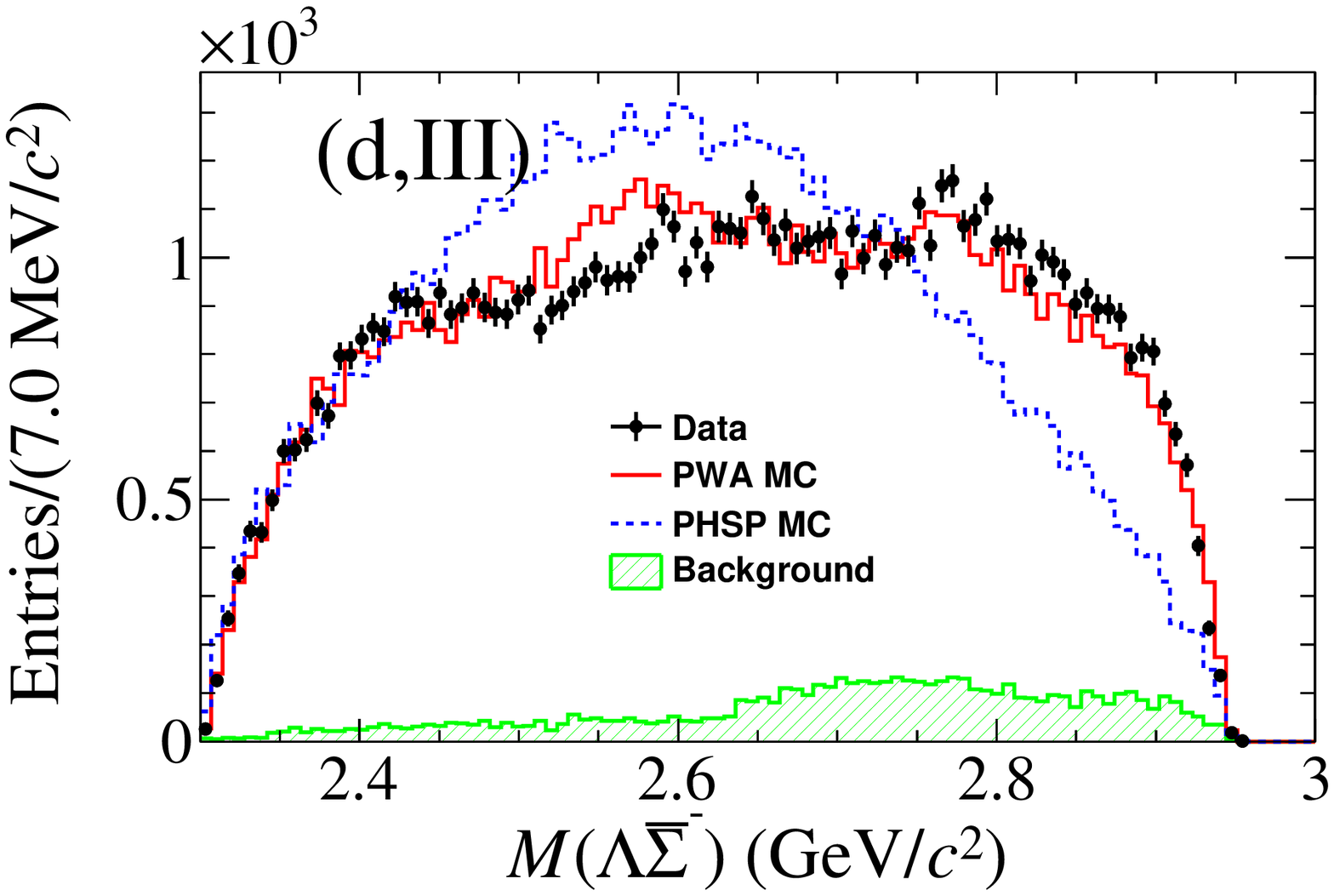}
        
        \caption{
            The distributions of $M(\bar{\Lambda}\pi)$~(I), $M(\Sigma\pi)$~(II), and $M(\bar{\Lambda}\Sigma)$~(III) from $J/\psi\rightarrow\bar{\Lambda}\pi^{+}\Sigma^{-}$ (a), $J/\psi\rightarrow\Lambda\pi^{-}\bar{\Sigma}^{+}$ (b), $J/\psi\rightarrow\bar{\Lambda}\pi^{-}\Sigma^{+}$ (c), and $J/\psi\rightarrow\Lambda\pi^{+}\bar{\Sigma}^{-}$ (d). The black dots with error bars are from data in the signal region. The green histograms are from data in the $\Sigma$ sideband regions. The red solid curves are PWA MC events normalized to the statistics of data. The blue solid curves represent PHSP MC events normalized to the statistics of data.
        }
        \label{fig:dalitzfit}
    \end{figure*}

    To determine the detection efficiency, the PWA is performed based on the TF-PWA framework~\cite{TFPWA:2022}, and the complex coupling constant of the amplitudes are determined using an unbinned maximum likelihood fit. In the log-likelihood calculation, since we use sideband events to describe the background in the signal region, the likelihood values of sideband events are given negative weights, and subtracted as $S=-(\ln L_{\rm{signal}}-\ln L_{\rm{sideband}})$. The $L_{\rm{signal}}$ is the likelihood values of events in the $\Sigma$ signal region of the data, and $L_{\rm{sideband}}$ is the likelihood values of events in the $\Sigma$ sideband region of data. The TF-PWA incorporates 11 potential intermediate excited states: $\Lambda(1380)$, $\Lambda(1405)$, $\Lambda(1520)$, $\Lambda(1600)$, $\Lambda(1690)$, $\Lambda(1820)$, $\Lambda(1890)$, $\Sigma(1385)$, $\Sigma(1660)$, $\Sigma(1670)$ and $\Sigma(1750)$. All of these resonances are described with relativistic Breit-Wigner functions, and their masses and widths are fixed to the world averages~\cite{PDG:2022}. To estimate the detection efficiency, the results from the PWA are used to generate MC events  with a better consistency to data (PWA MC events). For each decay mode, we generate $10^6$ MC events according to the PWA results. Figure~\ref{fig:dalitzfit} shows the distributions of $M(\bar{\Lambda}\pi)$, $M(\Sigma\pi)$, $M(\bar{\Lambda}\Sigma)$ from PWA MC samples, phase-space (PHSP) MC samples, signal and sideband regions of the data. The discrepancies between data and PWA MC samples are acceptable for the estimation of detection efficiency. In addition, the impact of these discrepancies is discussed in Sec.~\ref{sec:syst}.
    
    To compensate the discrepancy in $\Lambda$ reconstruction efficiency between the data and MC simulation, the PWA MC samples are weighted to match the reconstruction efficiency of data. The two-dimensional $\Lambda$ reconstruction efficiency within each $p$-$|\rm{cos\theta}|$ interval is studied using the control sample $\jpsi\go \bar{p}K^{+}\Lambda$, where $p$ is momentum and $\theta$ is the polar angle of $\Lambda$. A simultaneous fit is performed on the recoiling mass distribution of the $\bar{p}K^{+}$ system for events passing $\Lambda$ reconstruction and selection, events in the reconstructed $\Lambda$ sideband regions [$M(p\pi^{-})\in[1.096, 1.106]\cup[1.126, 1.136]~{\rm GeV}/c^2$], and for events failing to pass $\Lambda$ reconstruction or selection. A double Gaussian function is used to describe the signal shape and a second-order polynomial function is used to describe the background shape. The free parameters in the simultaneous fit are the signal event yields denoted as $n_{1}$, $n_{2}$, and $n_{3}$, respectively, for the aforementioned event samples. The efficiency of $\Lambda$ reconstruction is calculated by $\varepsilon^{j} = (n_{1}-n_{2})/(n_{1}+n_{3})$. The reconstruction efficiency of $\bar{\Lambda}$ is determined in a similar way to the $\jpsi\go pK^{-}\bar{\Lambda}$ control sample. The correction factors are calculated as $f^{j} = \varepsilon^{j}_{\rm data}/\varepsilon^{j}_{\rm MC}$, where $\varepsilon^{j}_{\rm data}$ and $\varepsilon^{j}_{\rm MC}$ are the reconstruction efficiencies of data and MC samples, respectively, and $j$ is the index of the $j$th $p$-$|\rm{cos\theta}|$ interval. As an example, the reconstruction efficiencies of 2019 $J/\psi$ data and MC samples are presented in Tables~\ref{tab:lambdaEff} and~\ref{tab:antilambdaEff}. The detection efficiency is determined by 
    \begin{equation}
	    \varepsilon_{\rm weighted} = \frac{\Sigma_{j}N^{j}_{\rm select}\times f^{j}}{\Sigma_{j}N^{j}_{\rm generate}},
    	\label{eq1}
    \end{equation}
    where $N^{j}_{\rm select}$ and $N^{j}_{\rm generate}$ are the number of MC events passing the event selection and the number of generated MC events in each interval.

    \begin{table*}[htbp]
    	\centering
    	\caption{The $\Lambda$ reconstruction efficiency in momentum and $\cos\theta$ intervals. Uncertainties are statistical only.}
    	\begin{tabular}{c|ccccc}\hline\hline
    		$\varepsilon_{\rm data}$~($\%$) & & & $p$~(GeV/$c$) & & \\ \hline
    		$|\cos\theta|$ & (0, 0.3) & (0.3, 0.5) & (0.5, 0.65) & (0.65, 0.75) & (0.75, 1.0) \\ \hline
    		(0.00, 0.20) & $8.88\pm0.21$ & $28.62\pm0.17$ & $33.05\pm0.17$ & $35.79\pm0.19$ & $37.63\pm0.21$ \\
    		(0.20, 0.40) & $8.82\pm0.22$ & $28.33\pm0.18$ & $31.75\pm0.18$ & $34.80\pm0.21$ & $37.09\pm0.22$ \\
    		(0.40, 0.65) & $7.65\pm0.18$ & $26.47\pm0.19$ & $29.63\pm0.17$ & $31.00\pm0.21$ & $32.68\pm0.22$ \\
    		(0.65, 1.00) & $4.39\pm0.11$ & $15.10\pm0.12$ & $19.63\pm0.16$ & $21.59\pm0.22$ & $24.18\pm0.28$ \\
    		\hline\hline
    		$\varepsilon_{\rm MC}$~($\%$) & & & $p$~(GeV/$c$) & & \\ \hline
    		$|\cos\theta|$ & (0, 0.3) & (0.3, 0.5) & (0.5, 0.65) & (0.65, 0.75) & (0.75, 1.0) \\ \hline
    		(0.00, 0.20) &$10.40\pm0.15$ & $29.54\pm0.12$ & $34.20\pm0.09$ & $36.11\pm0.12$ & $37.90\pm0.13$ \\
    		(0.20, 0.40) & $9.92\pm0.16$ & $28.77\pm0.12$ & $33.16\pm0.11$ & $35.02\pm0.13$ & $37.03\pm0.14$ \\
    		(0.40, 0.65) & $8.88\pm0.12$ & $26.83\pm0.11$ & $31.08\pm0.11$ & $32.88\pm0.13$ & $34.53\pm0.14$ \\
    		(0.65, 1.00) & $5.28\pm0.09$ & $15.73\pm0.06$ & $19.91\pm0.10$ & $21.39\pm0.13$ & $24.38\pm0.15$ \\
    		\hline\hline
    	\end{tabular}
    	\label{tab:lambdaEff}
    \end{table*}
    
    \begin{table*}[htbp]
    	\centering
    	\caption{The $\bar{\Lambda}$ reconstruction efficiency in momentum and $\cos\theta$ intervals. Uncertainties are statistical only.}
    	\begin{tabular}{c|ccccc}\hline\hline
    		$\varepsilon_{\rm data}$~($\%$) & & & $p$~(GeV/$c$) & & \\ \hline
    		$|\cos\theta|$ & (0, 0.3) & (0.3, 0.5) & (0.5, 0.65) & (0.65, 0.75) & (0.75, 1.0) \\ \hline
    		(0.00, 0.20) & $8.00\pm0.21$ & $27.00\pm0.18$ & $30.77\pm0.17$ & $33.56\pm0.20$ & $36.20\pm0.20$ \\
    		(0.20, 0.40) & $8.05\pm0.21$ & $26.37\pm0.18$ & $30.61\pm0.17$ & $32.77\pm0.20$ & $34.50\pm0.21$ \\
    		(0.40, 0.65) & $7.46\pm0.18$ & $25.00\pm0.16$ & $28.90\pm0.16$ & $30.51\pm0.21$ & $31.45\pm0.22$ \\
    		(0.65, 1.00) & $3.88\pm0.10$ & $14.15\pm0.12$ & $19.12\pm0.16$ & $21.53\pm0.20$ & $23.80\pm0.29$ \\
    		\hline\hline
    		$\varepsilon_{\rm MC}$~($\%$) & & & $p$~(GeV/$c$) & & \\ \hline
    		$|\cos\theta|$ & (0, 0.3) & (0.3, 0.5) & (0.5, 0.65) & (0.65, 0.75) & (0.75, 1.0) \\ \hline
    		(0.00, 0.20) & $8.84\pm0.14$ & $26.96\pm0.12$ & $31.96\pm0.11$ & $35.13\pm0.12$ & $37.52\pm0.13$ \\
    		(0.20, 0.40) & $8.46\pm0.14$ & $25.94\pm0.12$ & $30.82\pm0.11$ & $34.28\pm0.13$ & $36.23\pm0.13$ \\
    		(0.40, 0.65) & $8.01\pm0.12$ & $23.98\pm0.11$ & $28.37\pm0.10$ & $31.10\pm0.12$ & $33.65\pm0.14$ \\
    		(0.65, 1.00) & $4.37\pm0.07$ & $13.53\pm0.07$ & $18.33\pm0.09$ & $20.47\pm0.12$ & $23.91\pm0.14$ \\
    		\hline\hline
    	\end{tabular}
    	\label{tab:antilambdaEff}
    \end{table*}
    
    Similarly, the two-dimensional tracking and PID efficiencies of charged pions have been studied via the decay $J/\psi\rightarrow\pi^{+}\pi^{-}\pi^{0}$. The correction factors, calculated with Eq.~(\ref{eq1}), are used to weight the efficiency of $\pi^{\pm}$ from $J/\psi$, and the final corrected detection efficiency ($\varepsilon$) for each decay mode is shown in Table~\ref{tab:reweight_result}.
    
    \begin{table*}[htbp]
    	\caption{Efficiency changes with $\Lambda$ reconstruction efficiency correction and $\pi$ tracking and PID correction. The $f(\Lambda)$ and $f(\pi)$ are the correction factor of the $\Lambda$ reconstruction efficiency correction, and the $\pi$ tracking and PID correction, respectively. The $\varepsilon_{\rm ori}$ and $\varepsilon$ are the original and corrected detection efficiency, respectively.}
    	\centering
    	\begin{tabular}{ccccc}
    		\hline\hline
    		Decay mode & $\varepsilon_{\rm ori}$ & $f(\Lambda)$ & $f(\pi)$ & $\varepsilon$ \\ \hline
    		$J/\psi\rightarrow\bar{\Lambda}\pi^{+}\Sigma^{-}$ & $34.59\%$ & $0.9784$ & $0.9988$ & $33.80\%$ \\ 
    		$J/\psi\rightarrow\Lambda\pi^{-}\bar{\Sigma}^{+}$ & $36.34\%$ & $0.9612$ & $0.9989$ & $34.89\%$ \\ 
    		$J/\psi\rightarrow\bar{\Lambda}\pi^{-}\Sigma^{+}$ & $31.65\%$ & $0.9791$ & $0.9987$ & $30.95\%$ \\ 
    		$J/\psi\rightarrow\Lambda\pi^{+}\bar{\Sigma}^{-}$ & $34.18\%$ & $0.9624$ & $0.9986$ & $32.85\%$ \\ \hline\hline
    	\end{tabular}
    	\label{tab:reweight_result}
    \end{table*}

\section{\boldmath Branching fractions}
\label{sec:BR}

\begin{figure*}[htbp]
        \centering
        \includegraphics[width=2.8in]{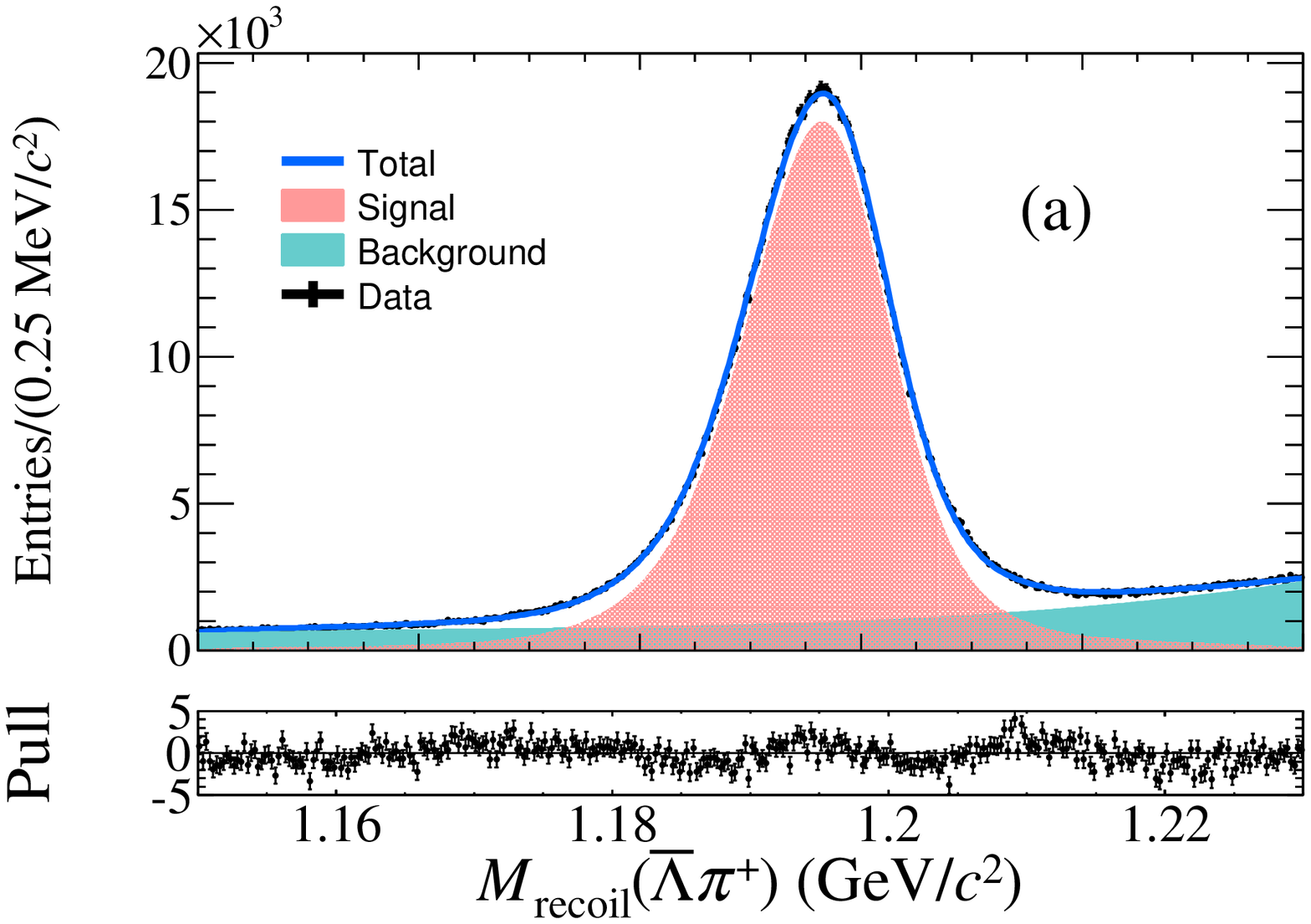}
        \includegraphics[width=2.8in]{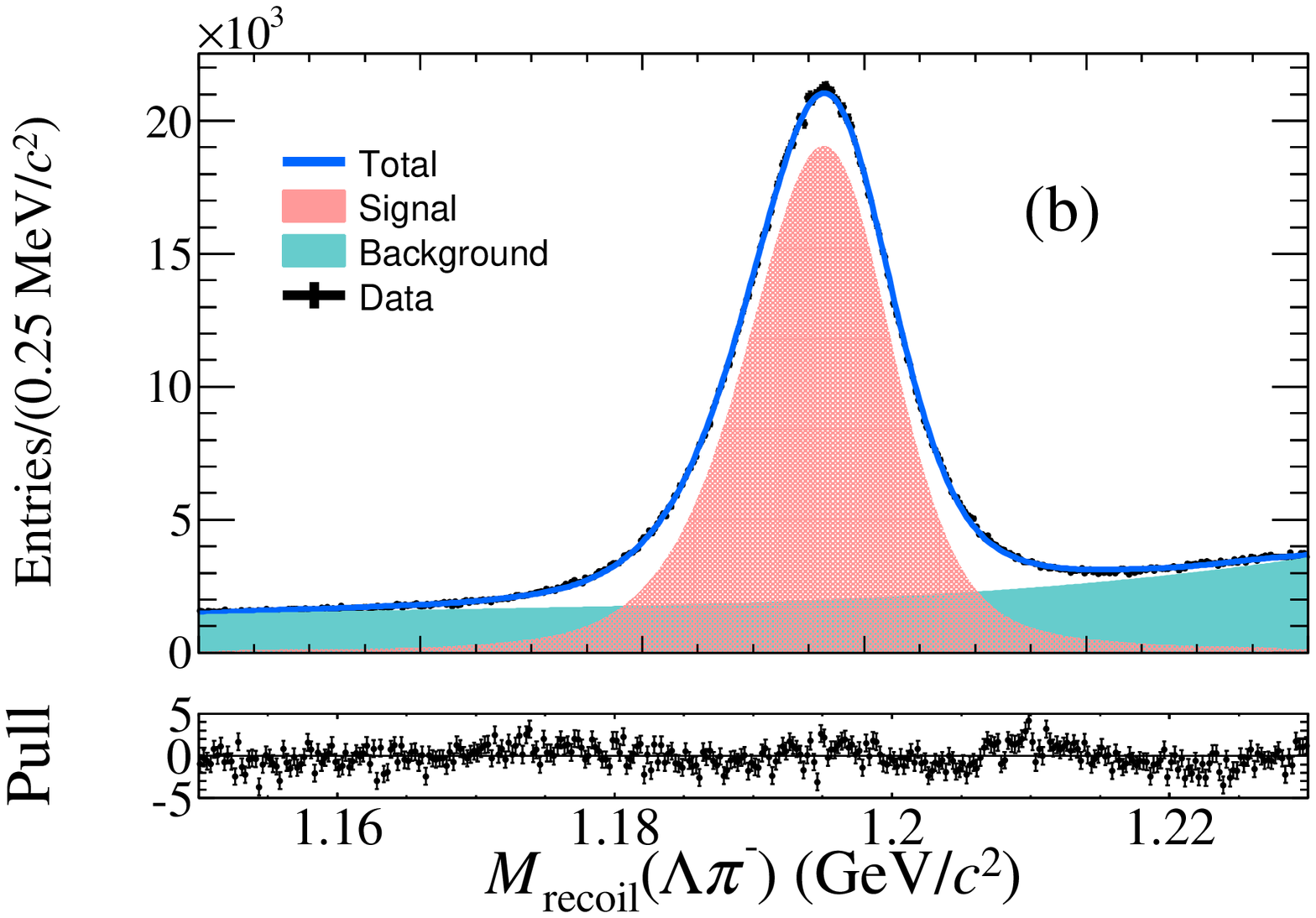}
        \includegraphics[width=2.8in]{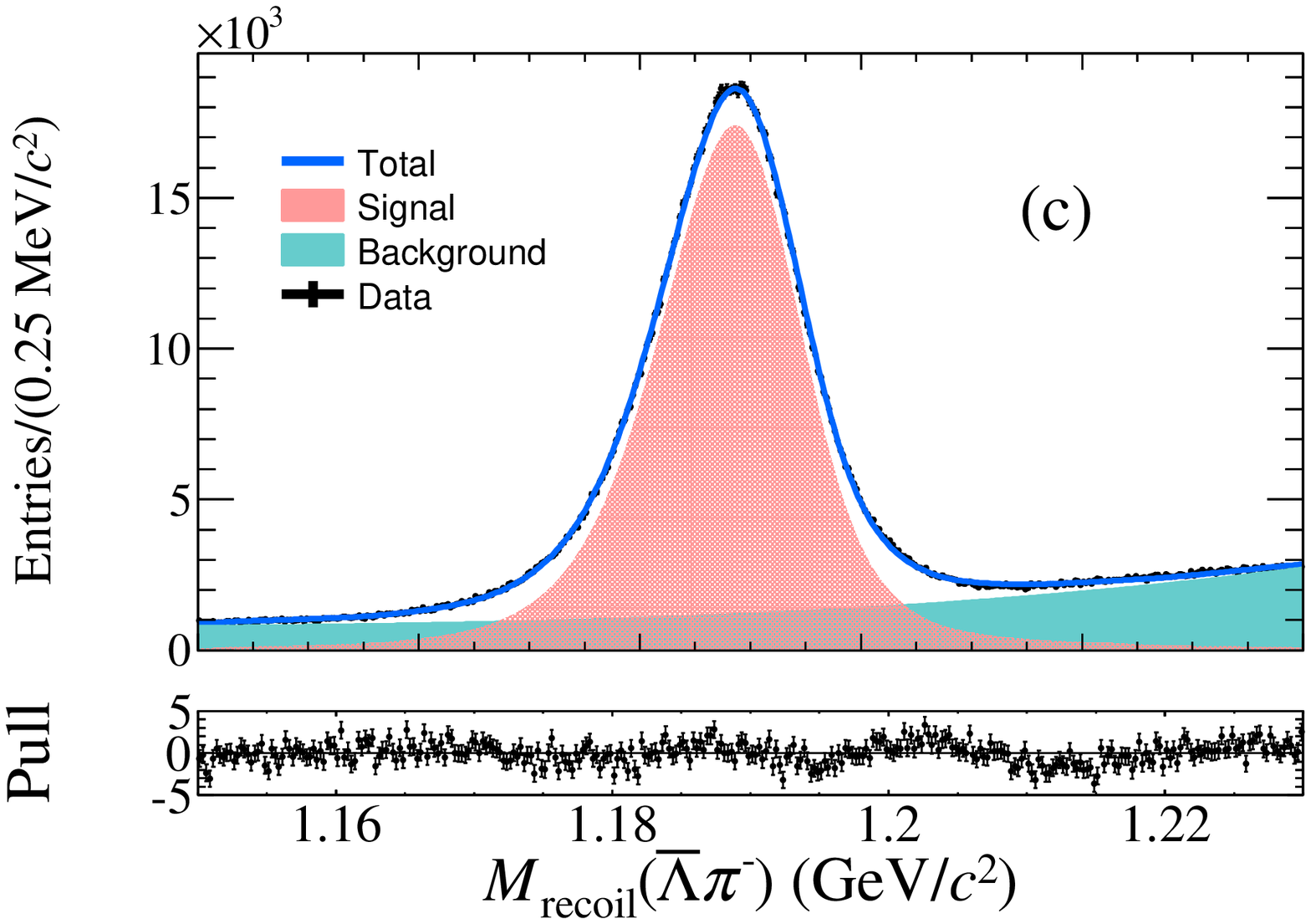}
        \includegraphics[width=2.8in]{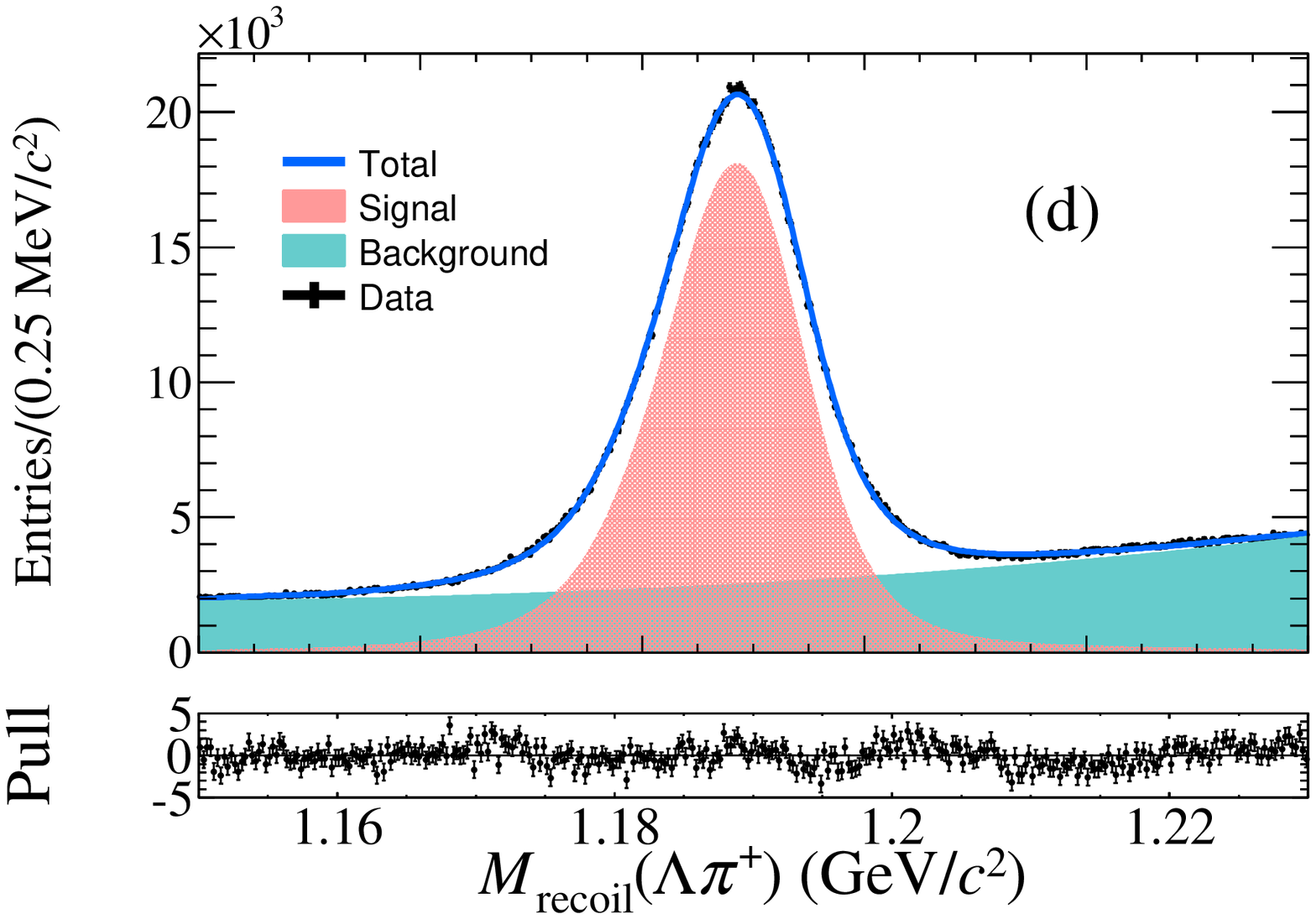}
        \caption{
            The fit results of the $M_{\rm recoil}(\bar{\Lambda}\pi)$ distributions from $J/\psi\rightarrow\bar{\Lambda}\pi^{+}\Sigma^{-}$ (a), $J/\psi\rightarrow\Lambda\pi^{-}\bar{\Sigma}^{+}$ (b), $J/\psi\rightarrow\bar{\Lambda}\pi^{-}\Sigma^{+}$ (c), and $J/\psi\rightarrow\Lambda\pi^{+}\bar{\Sigma}^{-}$ (d). The black dots with error bars are data, the blue curves are the total fits, the red curves are the signals, and the green curves are the backgrounds.
        }
        \label{fig:fit}
    \end{figure*}

    A binned maximum likelihood fit is performed to the recoiling mass distribution of the $\bar{\Lambda}\pi$ system to determine the number of signal events, as shown in Fig.~\ref{fig:fit}. The signal is described by a RooKeysPdf~\cite{Cranmer:2001} of the signal MC events convolved with a Gaussian function to account for the difference of resolution between data and MC simulation, and the background is described by a third-order Chebyshev function.
    
\begin{figure*}[htbp]
        \centering
        \includegraphics[width=2.5in]{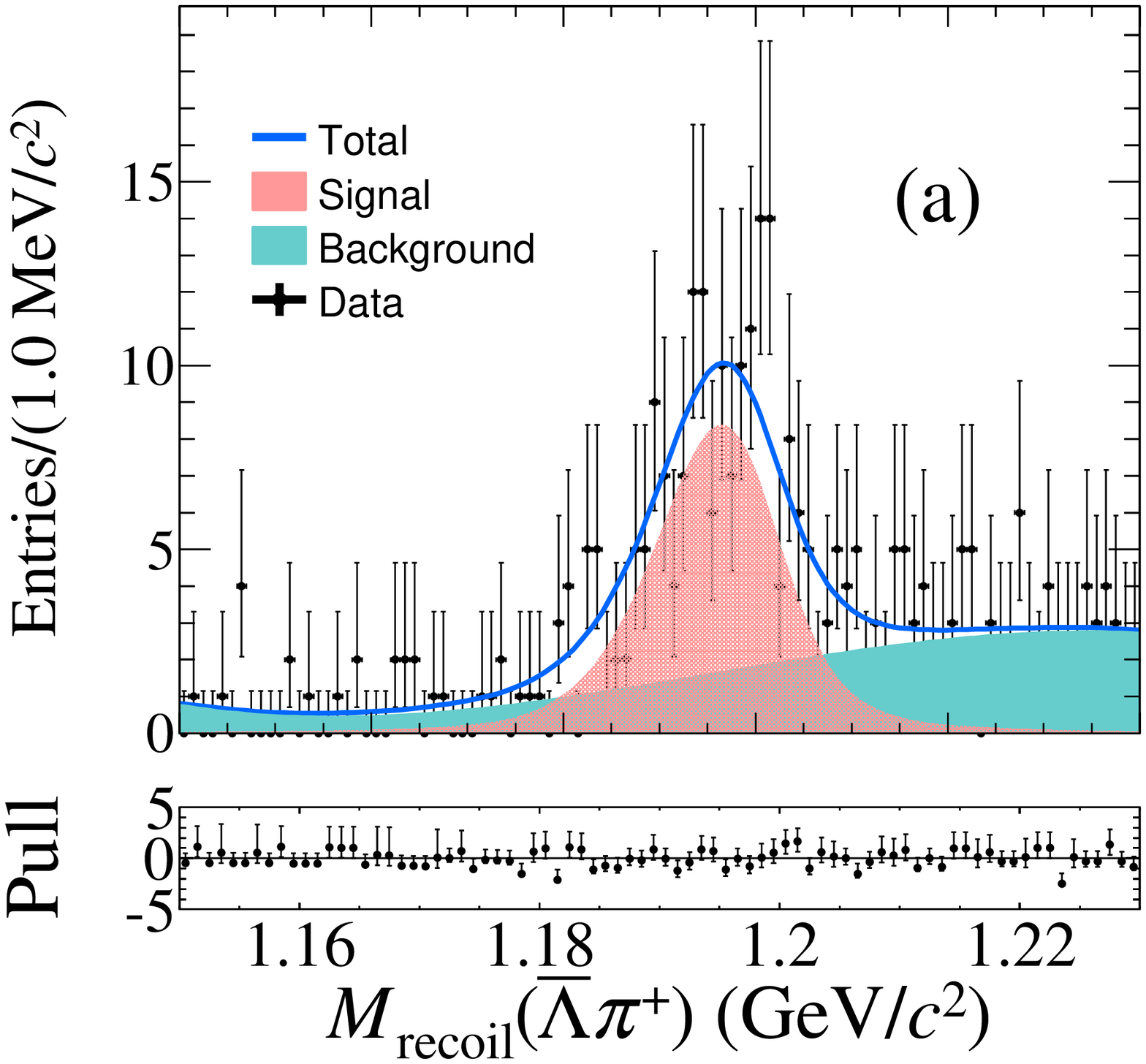}
        \includegraphics[width=2.5in]{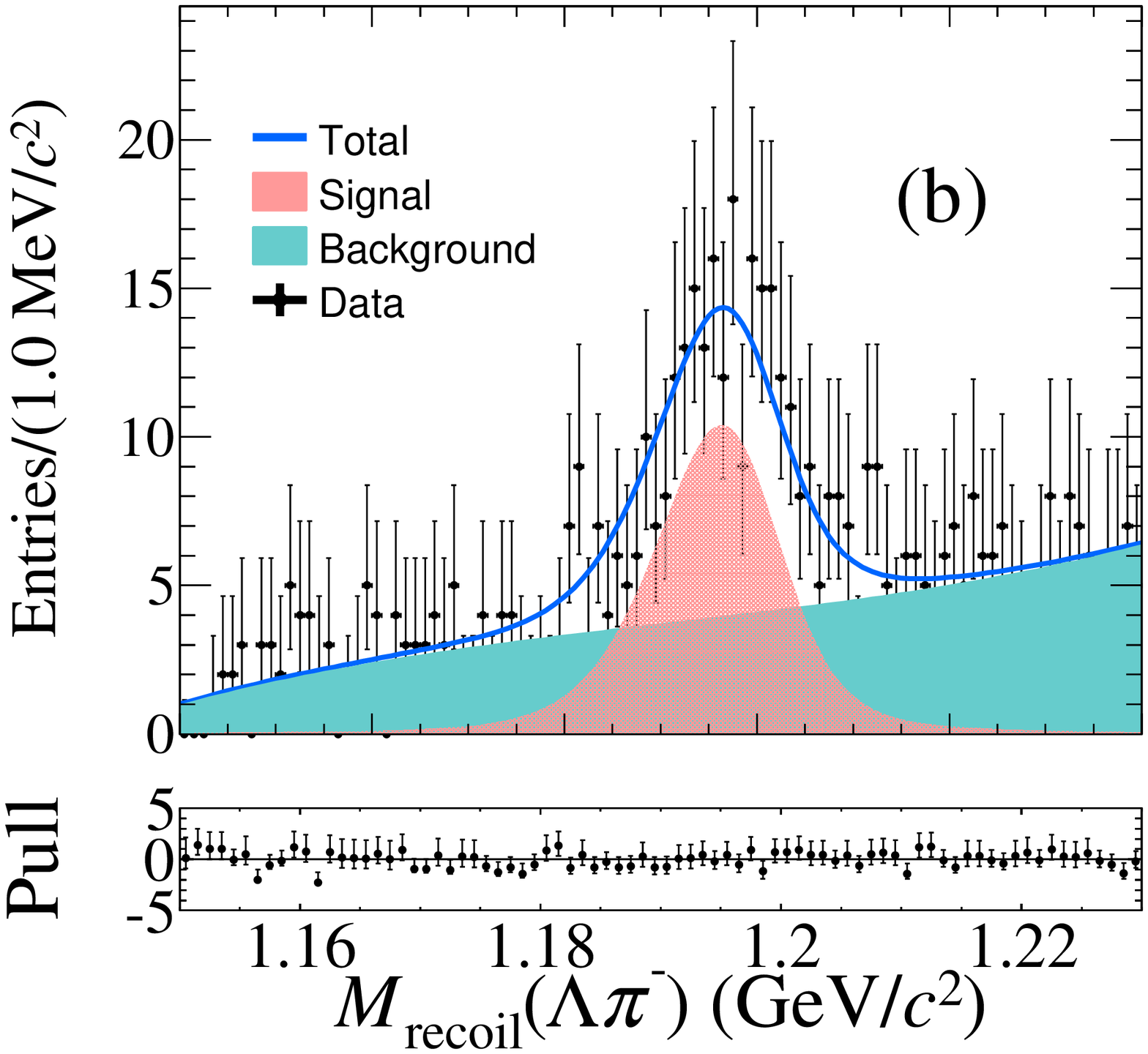}
        \includegraphics[width=2.5in]{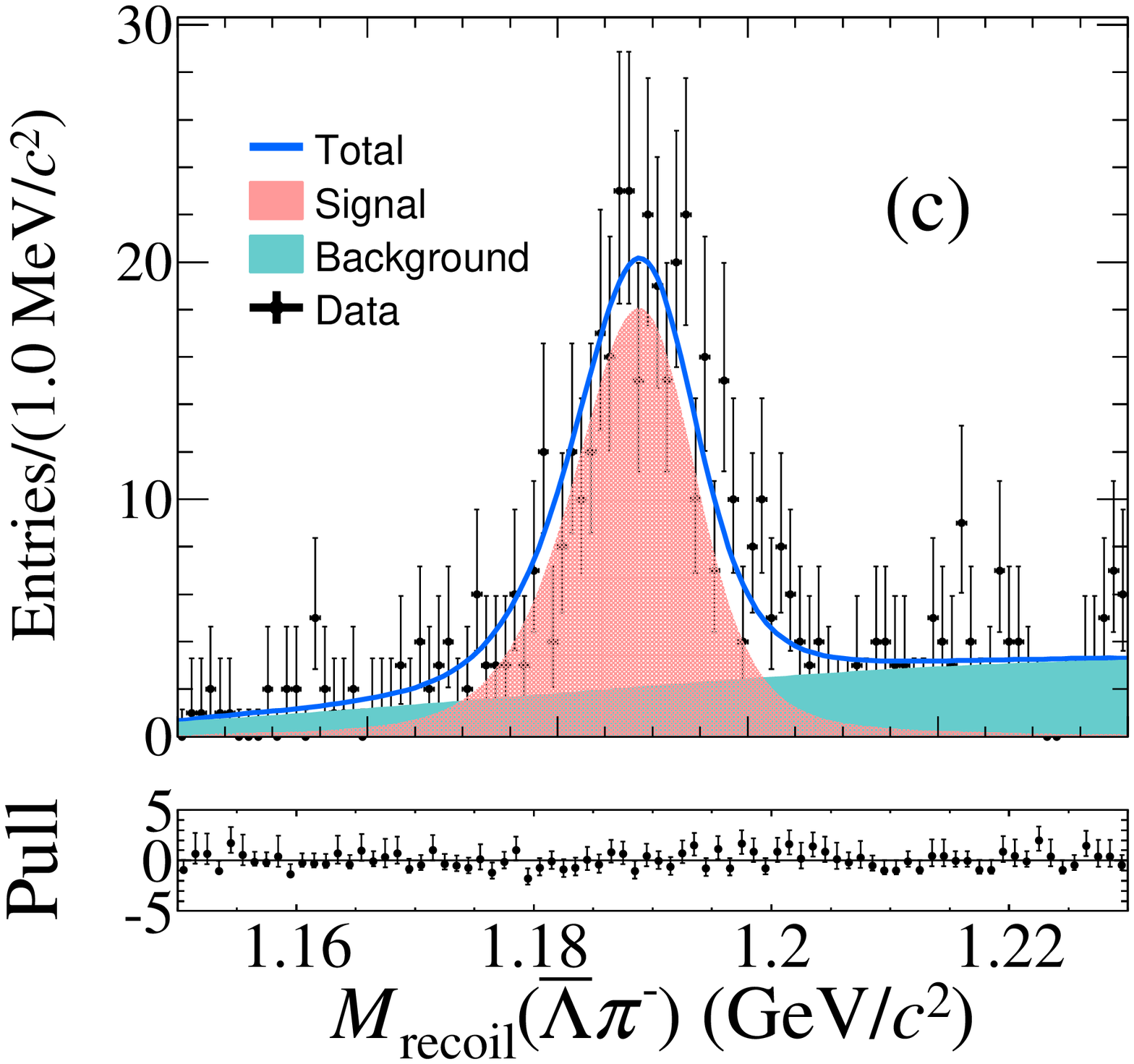}
        \includegraphics[width=2.5in]{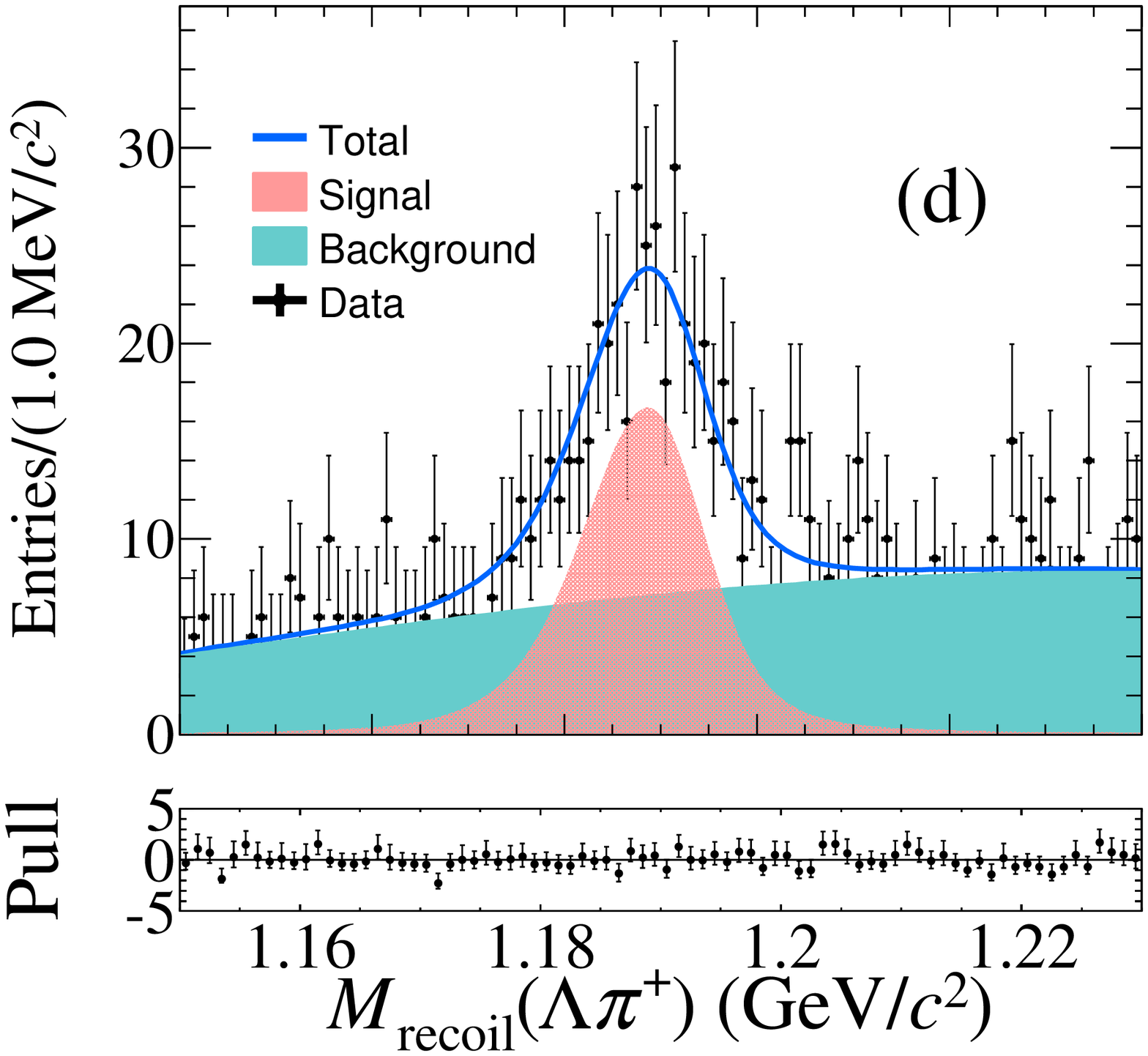}
        \caption{
            The fit results of the $M_{\rm recoil}(\bar{\Lambda}\pi)$ distributions from $\ee\go\lpsm$~(a), $\ee\go\lpsap$~(b), $\ee\go\lpsp$~(c), and $\ee\go\lpsam$~(d). The black dots with error bars are data taken at $\sqrt{s}=3.080~{\rm GeV}$, the blue curves are the total fits, the red curves are the signals, and the green curves are the backgrounds.
        }
        \label{fig:QEDfit}
    \end{figure*}

    An unbinned maximum likelihood fit process is performed on data taken at $\sqrt{s}=3.080~{\rm GeV}$, where the parameters of the Gaussian resolution function are fixed to those obtained in $J/\psi$ data. The continuum events are then normalized to the $\jpsi$ data after taking into account the luminosity difference. Figure~\ref{fig:QEDfit} shows the fit result of the continuum events.

    The branching fraction (${\cal B}$) is obtained via
    \begin{equation}
        \begin{split}
	    {\cal B} = \frac{N_{\rm obs} - N_{\rm QED}}{N_{J/\psi} \cdot {\cal B}_{\Lambda\go p\pi^{-}} \cdot \varepsilon },
        \end{split}
	\label{eq2}
    \end{equation}
    where $N_{\rm obs}$ is the number of signal events obtained from the fit, $N_{J/\psi}$ is the number of $\jpsi$ events in data~\cite{HXYang:2022}, ${\cal B}_{\Lambda\go p\pi^{-}}$ is the branching fraction of $\Lambda\go p\pi^{-}$~\cite{PDG:2022}, and $\varepsilon$ is the corrected detection efficiency. $N_{\rm QED}$ is the number of QED background events estimated with data taken at $\sqrt{s}=3.080~{\rm GeV}$ after considering the difference of luminosity and energy-dependent cross section, obtained via
    \begin{equation}
	    N_{\rm QED} = N^{\rm fit}_{3.080} \times \frac{{\cal L}_{\jpsi}}{{\cal L}_{\rm 3.080}} \times \frac{3.080^{2}}{3.097^{2}},
    	\label{eq3}
    \end{equation}
    where $N^{\rm fit}_{3.080}$ is the number of signal events from the fit, the ${\cal L}_{\jpsi}=3083~{\rm pb}^{-1}$ and ${\cal L}_{\rm 3.080}=166~{\rm pb}^{-1}$ are the integrated luminosities for $\jpsi$ and $\sqrt{s}=3.080$ GeV data, respectively~\cite{HXYang:2022}. The results are summarized in Table~\ref{tab:BR}.

    \begin{table*}[htp]
	\centering
	\caption{Signal yields in $\jpsi$ data~($N_{\rm obs}$), QED background yields~[$N_{\rm QED}$ from Eq.~(\ref{eq3})], branching fractions obtained in this work~[${\cal B}$ from Eq.~(\ref{eq2})] and previous results from Table~\ref{tab:Old} ($\mathcal B_{\rm old}$). The first uncertainties are statistical and the second ones are systematic. The $\Sigma$+$\bar{\Sigma}$ stands for the sum of the charge-conjugate channels.}
	\begin{tabular}{ccccc}
		\hline\hline
		Decay mode & $N_{\rm obs}$ & $N_{\rm QED}$ & ${\cal B}~(\times10^{-4})$ & $\mathcal B_{\rm old}$~($\times10^{-4}$) \\ \hline
		$\bar{\Lambda}\pi^{+}\Sigma^{-}$ & $1,322,142\pm1,773$ & $2,805\pm358$ & $6.06\pm0.01\pm0.21$ & - \\
		$\Lambda\pi^{-}\bar{\Sigma}^{+}$ & $1,387,438\pm1,986$ & $3,422\pm450$ & $6.15\pm0.01\pm0.20$ & - \\
		$\Sigma^{-}+\bar{\Sigma}^{+}$ & - & - & $12.21\pm0.02\pm0.38$ & $9.9\pm1.6$ \\ \hline
		$\bar{\Lambda}\pi^{-}\Sigma^{+}$ & $1,257,698\pm1,679$ & $5,946\pm463$ & $6.27\pm0.01\pm0.23$ & - \\
		$\Lambda\pi^{+}\bar{\Sigma}^{-}$ & $1,310,437\pm1,931$ & $5,507\pm564$ & $6.16\pm0.01\pm0.23$ & - \\ 
		$\Sigma^{+}+\bar{\Sigma}^{-}$ & - & - & $12.44\pm0.02\pm0.45$ & $13.5\pm1.3$ \\
		\hline\hline
	\end{tabular}
	\label{tab:BR}
\end{table*}

\section{\boldmath Systematic Uncertainties}
\label{sec:syst}

    The sources of systematic uncertainties in the branching fraction measurements are summarized in Table~\ref{tab:Syst}, and $\Sigma$+$\bar{\Sigma}$ stands for the sum of the charge-conjugate channels.
    
    \begin{table*}[htp]
    	\centering
    	\caption{Relative systematic uncertainties in the branching fraction measurements (in $\%$).}
    	\begin{tabular}{lcccccc}
    	    \hline\hline
                Source & $\bar{\Lambda}\pi^{+}\Sigma^{-}$ & $\Lambda\pi^{-}\bar{\Sigma}^{+}$ & $\Sigma^{-}$+$\bar{\Sigma}^{+}$ & $\bar{\Lambda}\pi^{-}\Sigma^{+}$ & $\Lambda\pi^{+}\bar{\Sigma}^{-}$ &
                $\Sigma^{+}$+$\bar{\Sigma}^{-}$ \\
                \hline
                Number of $J/\psi$ events & $0.44$ & $0.44$ & $0.44$ & $0.44$ & $0.44$ & $0.44$ \\
                $\Lambda\rightarrow p\pi^{-}$ & $0.78$ & $0.78$ & $0.78$ & $0.78$ & $0.78$ & $0.78$ \\
                $\Lambda$ reconstruction & $0.76$ & $0.78$ & $0.55$ & $0.74$ & $0.74$ & $0.53$ \\
                Tracking and PID for $\pi$ & $0.09$ & $0.09$ & $0.07$ & $0.10$ & $0.10$ & $0.08$ \\
                MC statistics      & $0.15$ & $0.15$ & $0.10$ & $0.15$ & $0.15$ & $0.10$ \\
                QED background     & $2.06$ & $2.24$ & $2.15$ & $2.92$ & $2.85$ & $2.89$ \\
                Fit method         & $1.50$ & $1.63$ & $1.11$ & $0.75$ & $1.32$ & $0.76$ \\
                PWA (intermediate resonances) & $1.00$ & $1.00$ & $1.00$ & $1.00$ & $1.00$ & $1.00$ \\
                PWA (sideband region)         & $0.14$ & $0.14$ & $0.14$ & $0.04$ & $0.12$ & $0.08$ \\
                Signal shape       & $1.75$ & $0.75$ & $1.25$ & $1.57$ & $1.13$ & $1.35$ \\
                \hline
                Total      & $3.46$ & $3.27$ & $3.09$ & $3.73$ & $3.68$ & $3.58$ \\
                \hline\hline
    	\end{tabular}
    	\label{tab:Syst}
    \end{table*}
    
    The uncertainty due to the number of $J/\psi$ events is $0.44\%$~\cite{HXYang:2022}, and the systematic uncertainty on the branching fraction of the decay $\Lambda\rightarrow p\pi^{-}$ is $0.78\%$, quoted from PDG~\cite{PDG:2022}.
    
    The uncertainty due to the $\Lambda$ reconstruction efficiency correction is estimated with the uncertainties of the correction factor $f^{j}$, which propagates to the detection efficiency according to
    
    \begin{equation}
        \begin{split}
    	\frac{\delta\varepsilon_{\rm weighted}}{\varepsilon_{\rm weighted}} = \frac{ \sqrt{ \Sigma_{j} (N^{j}_{\rm select}\times\delta f^{j})^{2} } }{\Sigma_{j}N^{j}_{\rm generate} \times \varepsilon_{\rm weighted}}, \\
    	~\frac{\delta f^{j}}{f^{j}} = \sqrt{(\frac{\delta\varepsilon^{j}_{\rm data}}{\varepsilon^{j}_{\rm data}})^2+(\frac{\delta\varepsilon^{j}_{\rm MC}}{\varepsilon^{j}_{\rm MC}})^2},
    	\end{split}
    	\label{eq:errorprop}
    \end{equation}
    where $\delta\varepsilon_{\rm weighted}$, $\delta\varepsilon^{j}_{\rm data}$ and $\delta\varepsilon^{j}_{\rm MC}$ are the uncertainties of $\varepsilon_{\rm weighted}$, $\varepsilon^{j}_{\rm data}$ and $\varepsilon^{j}_{\rm MC}$ from Eq.~(\ref{eq1}), respectively. The reconstruction method of control samples is the same as data. The uncertainties due to the requirements of the $\Lambda$ mass window and the ratio of the decay length to the error of $\Lambda$ are considered. The uncertainties in the $\Lambda$ reconstruction are estimated to be $0.76\%$ for $J/\psi\rightarrow\bar{\Lambda}\pi^{+}\Sigma^{-}$, $0.78\%$ for $J/\psi\rightarrow\Lambda\pi^{-}\bar{\Sigma}^{+}$, and $0.74\%$ for $J/\psi\rightarrow\bar{\Lambda}\pi^{-}\Sigma^{+}$, and $J/\psi\rightarrow\Lambda\pi^{+}\bar{\Sigma}^{-}$.
    
    Similar to the method used for the $\Lambda$ reconstruction efficiency correction, the uncertainties due to the $\pi$ tracking and PID efficiency correction are estimated to be $0.09\%$ for $J/\psi\rightarrow\bar{\Lambda}\pi^{+}\Sigma^{-}$ and $J/\psi\rightarrow\Lambda\pi^{-}\bar{\Sigma}^{+}$, and $0.10\%$ for $J/\psi\rightarrow\bar{\Lambda}\pi^{-}\Sigma^{+}$ and $J/\psi\rightarrow\Lambda\pi^{+}\bar{\Sigma}^{-}$.
    
    After considering the effect of possible intermediate states, $10^6$ MC events are generated for each decay mode to estimate the relative uncertainty due to the limited MC statistics. The uncertainty in each decay mode is estimated to be $0.15\%$, which is calculated from $\sqrt{\frac{1-\varepsilon}{N\cdot\varepsilon}}$, where $\varepsilon$ is the detection efficiency and $N$ is the number of produced signal MC events. For the sum of charge-conjugate channels, MC statistics are added to calculate the statistical uncertainties. Then the systematic uncertainty of the sum of the two channels is $0.10\%$ for $J/\psi\rightarrow\bar{\Lambda}\pi^{+}\Sigma^{-}+c.c.$ and $J/\psi\rightarrow\bar{\Lambda}\pi^{-}\Sigma^{+}+c.c.$.
    
    The interference between the continuum processes and the $J/\psi$ decay amplitudes may affect the branching fraction calculation~\cite{YPGuo:2022}. The total cross section of $e^{+}e^{-}\rightarrow f$ at the $J/\psi$ peak can be written as
    \begin{equation}
        \begin{split}
        \sigma 
        & = \left |a_c + e^{i\phi}a_{J/\psi}\right |^{2} \\
        & = \left |\sqrt{\sigma_c} + e^{i\phi}\frac{\sqrt{12\pi\Gamma_{ee}\cdot\Gamma\cdot{\cal B}}}{im\Gamma}\right |^{2} \\ 
        & = \left |\sqrt{\sigma_c} + e^{i(\phi-\frac{\pi}{2})}\frac{\sqrt{12\pi{\cal B}_{ee}\cdot{\cal B}}}{m}\right |^{2},
        \end{split}
    	\label{eq6}
    \end{equation}
    \noindent where $m$, $\Gamma$, $\Gamma_{ee}$ and ${\cal B}_{ee}$ are the mass of $J/\psi$, the total width of $J/\psi$, the width of $J/\psi\rightarrow e^{+}e^{-}$, and the branching fraction of $J/\psi\rightarrow e^{+}e^{-}$, respectively~\cite{PDG:2022}. The ${\cal B}$, $a_{c}$, $a_{J/\psi}$, $\sigma_c$ and $\phi$ are our measured branching fraction, the amplitude of the continuum process, the amplitude of the resonance process, the cross section of the continuum process, and the relative phase between the amplitudes of the resonance and continuum processes, respectively. The measured branching fractions are a result of a combination of contributions from both the continuum and the resonance processes, and their differentiation is not feasible unless the value of $\phi$ is determined. Since we do not know $\phi$ from our data, we set $\sin{\phi}$ as a random function with a range of $[-1, 1]$ to estimate the interference effect. The statistical uncertainty of $N_{\rm QED}$ also affects the cross section of the continuum process and the interference between the continuum and the resonance processes. We model the fitted $N_{\rm QED}$ as a Gaussian function, whose mean and deviation are the central value and statistical uncertainty of the fitted $N_{\rm QED}$. To obtain the branching fraction distributions, we generate 10,000 random samples. The largest difference between the sampled branching fraction and the nominal branching fraction is taken as the uncertainty of the QED background, which is $2.06\%$ for $J/\psi\rightarrow\bar{\Lambda}\pi^{+}\Sigma^{-}$, $2.24\%$ for $J/\psi\rightarrow\Lambda\pi^{-}\bar{\Sigma}^{+}$, $2.92\%$ for $J/\psi\rightarrow\bar{\Lambda}\pi^{-}\Sigma^{+}$, and $2.85\%$ for $J/\psi\rightarrow\Lambda\pi^{+}\bar{\Sigma}^{-}$. The nominal treatments for the sum of charge-conjugate channels are identical, and their potential deviations from the truth exhibit a similar pattern. By adding their absolute systematic uncertainties and converting them to relative uncertainty, we can determine the overall systematic uncertainty for the sum of the two channels, which is $2.15\%$ for $J/\psi\rightarrow\bar{\Lambda}\pi^{+}\Sigma^{-}+c.c.$, and $2.89\%$ for $J/\psi\rightarrow\bar{\Lambda}\pi^{-}\Sigma^{+}+c.c.$. We include these as one source of systematic uncertainties due to interference between the continuum and $J/\psi$ decay amplitudes.
    
    The systematic uncertainty of the fitting originates from the fit range and the choice of the  background functions. The uncertainty due to the fit range is estimated by varying the range by $\pm 10~{\rm MeV}/c^2$. The fourth-order Chebyshev polynomial function is used to replace the background shape instead of the third-order Chebyshev polynomial function. The resulting largest difference
    to the original branching fraction is assigned as the systematic uncertainty, which is $1.50\%$ for $J/\psi\rightarrow\bar{\Lambda}\pi^{+}\Sigma^{-}$, $1.63\%$ for $J/\psi\rightarrow\Lambda\pi^{-}\bar{\Sigma}^{+}$, $0.75\%$ for $J/\psi\rightarrow\bar{\Lambda}\pi^{-}\Sigma^{+}$, and $1.32\%$ for $J/\psi\rightarrow\Lambda\pi^{+}\bar{\Sigma}^{-}$.
    
    In the PWA process, the selection of intermediate resonances and background events causes systematic uncertainty. On the basis of our solutions, additional resonances, $\Lambda$(1670) and $\Lambda$(1830),  for $J/\psi\rightarrow\bar{\Lambda}\pi^{+}\Sigma^{-} + c.c.$ and $J/\psi\rightarrow\bar{\Lambda}\pi^{-}\Sigma^{+} + c.c.$ are added to perform the PWA. Since the resonance parameters of intermediates in the PWA could be different from the real, the MC samples are regenerated with varying the masses and widths. The detection efficiency is calculated and compared using different MC samples. The comparison reveals a discrepancy of approximately $3\%$ in the detection efficiency between the PHSP MC sample and the PWA MC sample, despite the significant differences observed in the Dalitz plots, as shown in Fig.~\ref{fig:dalitzfit}. In contrast, the disparity between data and PWA MC is considerably smaller, with a calculated difference in detection efficiency among the PWA MC sample at the 1\% level. Consequently, we give an educated estimation of the systematic uncertainties of about 1\% due to intermediate resonances in the PWA. To estimate the systematic uncertainty of the background selection in the PWA, the modified sideband regions of $M(n\pi^{-})\in[1.15, 1.16]\cup[1.23, 1.24]~{\rm GeV}/c^2$ for $J/\psi\rightarrow\bar{\Lambda}\pi^{+}\Sigma^{-} + c.c.$, and $M(n\pi^{+})\in[1.142, 1.152]\cup[1.222, 1.232]~{\rm GeV}/c^2$ for $J/\psi\rightarrow\bar{\Lambda}\pi^{-}\Sigma^{+} + c.c.$ are chosen. The resulting difference in the detection efficiencies is assigned as the systematic uncertainty, which is $0.14\%$ for $J/\psi\rightarrow\bar{\Lambda}\pi^{+}\Sigma^{-}$, $0.14\%$ for $J/\psi\rightarrow\Lambda\pi^{-}\bar{\Sigma}^{+}$, $0.04\%$ for $J/\psi\rightarrow\bar{\Lambda}\pi^{-}\Sigma^{+}$, and $0.12\%$ for $J/\psi\rightarrow\Lambda\pi^{+}\bar{\Sigma}^{-}$, respectively. By applying the same method used to determine the systematic uncertainty due to QED background, the charge-conjugate channels are determined to be $0.14\%$ for $J/\psi\rightarrow\bar{\Lambda}\pi^{+}\Sigma^{-}+c.c.$, and $0.08\%$ for $J/\psi\rightarrow\bar{\Lambda}\pi^{-}\Sigma^{+}+c.c.$.
    
    The uncertainty from signal shape is caused by the utilization of different functions to describe the signal shape between the data and MC samples, especially for large statistics. The alternative fits with the triple-Gaussian function are used to describe the signal shape, and the differences between them are taken as the systematic uncertainty due to the parametrization of the signal shape, which are $1.75\%$ for $J/\psi\rightarrow\bar{\Lambda}\pi^{+}\Sigma^{-}$, $0.75\%$ for $J/\psi\rightarrow\Lambda\pi^{-}\bar{\Sigma}^{+}$, $1.57\%$ for $J/\psi\rightarrow\bar{\Lambda}\pi^{-}\Sigma^{+}$, and $1.13\%$ for $J/\psi\rightarrow\Lambda\pi^{+}\bar{\Sigma}^{-}$, respectively. For the sum of charge conjugate channels, their systematic uncertainties are correlated. By applying the same method used to determine the systematic uncertainty due to QED background, the charge-conjugate channels are determined to be $1.25\%$ for $J/\psi\rightarrow\bar{\Lambda}\pi^{+}\Sigma^{-}+c.c.$, and $1.35\%$ for $J/\psi\rightarrow\bar{\Lambda}\pi^{-}\Sigma^{+}+c.c.$.
    
    The total absolute systematic uncertainty in the branching fraction measurements is calculated by assuming the individual components to be independent, and adding their magnitudes in quadrature.

\section{\boldmath Summary and Discussions}

    Based on $(10087\pm44)\times10^6$ $\jpsi$ events collected with the BESIII detector, the branching fraction processes of $\jpsi\go\lpsm\cc$ and $\jpsi\go\lpsp\cc$ are measured to be consistent and 3 times more precise than the previous result. The branching fractions are summarized in Table~\ref{tab:BR} and compared with previous results.

    Table~\ref{tab:isospin} shows the test of the isospin symmetry in the two modes measured in this analysis. The common uncertainties due to the number of $J/\psi$ events and the branching fraction of the decay $\bar{\Lambda}\go\bar{p} \pi^{+}$ cancel out in the calculation of ratios. The ratios agree with one another within uncertainties. This indicates that these final states are produced dominantly by strong decays of the $\jpsi$, and the electromagnetic decay amplitude is small~\cite{MoXH:2022}. This is confirmed by the cross sections of continuum production of these modes measured with the continuum data taken at $\sqrt{s}=3.080~{\rm GeV}$ as shown in Table~\ref{tab:QED}.

    \begin{table*}[htp]
    	\centering
    	\caption{Isospin symmetry test.}
    	\begin{tabular}{cc}\hline\hline
    		 ${\cal B}(J/\psi\rightarrow\bar{\Lambda}\pi^{+}\Sigma^{-})/{\cal B}(J/\psi\rightarrow\bar{\Lambda}\pi^{-}\Sigma^{+})$ & $0.97\pm0.05$ \\
    		 ${\cal B}(J/\psi\rightarrow\Lambda\pi^{-}\bar{\Sigma}^{+})/{\cal B}(J/\psi\rightarrow\Lambda\pi^{+}\bar{\Sigma}^{-})$ & $1.00\pm0.05$ \\
    		 ${\cal B}(J/\psi\rightarrow\bar{\Lambda}\pi^{+}\Sigma^{-}+c.c.)/{\cal B}(J/\psi\rightarrow\bar{\Lambda}\pi^{-}\Sigma^{+}+c.c.)$ & $0.98\pm0.04$ \\
    		\hline\hline
    	\end{tabular}
    	\label{tab:isospin}
    \end{table*}

    \begin{table*}[htp]
    	\caption{The continuum cross sections of $\ee\go\lps$ systems at $\sqrt{s}=3.080~{\rm GeV}$. Uncertainties are statistical only.}
    	\centering
    	\begin{tabular}{cc}
    		\hline\hline
			Decay mode & ${\sigma}_{\rm QED}~({\rm pb})$ \\ \hline
			$\ee\go\lpsm$  & $4.21\pm0.54$ \\ 
			$\ee\go\lpsap$ & $4.98\pm0.65$ \\ 
			$\ee\go\lpsp$  & $9.75\pm0.76$ \\ 
			$\ee\go\lpsam$ & $8.51\pm0.87$ \\ 
			\hline\hline
    	\end{tabular}
    	\label{tab:QED}
    \end{table*}
	
    With the branching fractions ${\cal B}[\psi(3686)\to\bar{\Lambda}\pi^{+}\Sigma^{-}+c.c.]$ and ${\cal B}[\psi(3686)\to\bar{\Lambda}\pi^{-}\Sigma^{+}+c.c.]$ measured in Ref.~\cite{BR:2013} and assuming that the measurements of $\jpsi$ decays and $\psi(3686)$ decays are completely independent, the ratios to the corresponding $\jpsi$ decays are calculated to be $(11.5\pm1.4)\%$ and $(12.4\pm1.4)\%$, respectively, in good agreement with the ``$12\%$ rule". The precision of the ratios is limited mainly by the uncertainties of $\psi(3686)$ decay branching fractions, and measurements with a larger $\psi(3686)$ data sample will improve them.
    
\begin{figure*}[htbp]
        \centering
        \includegraphics[width=2.0in]{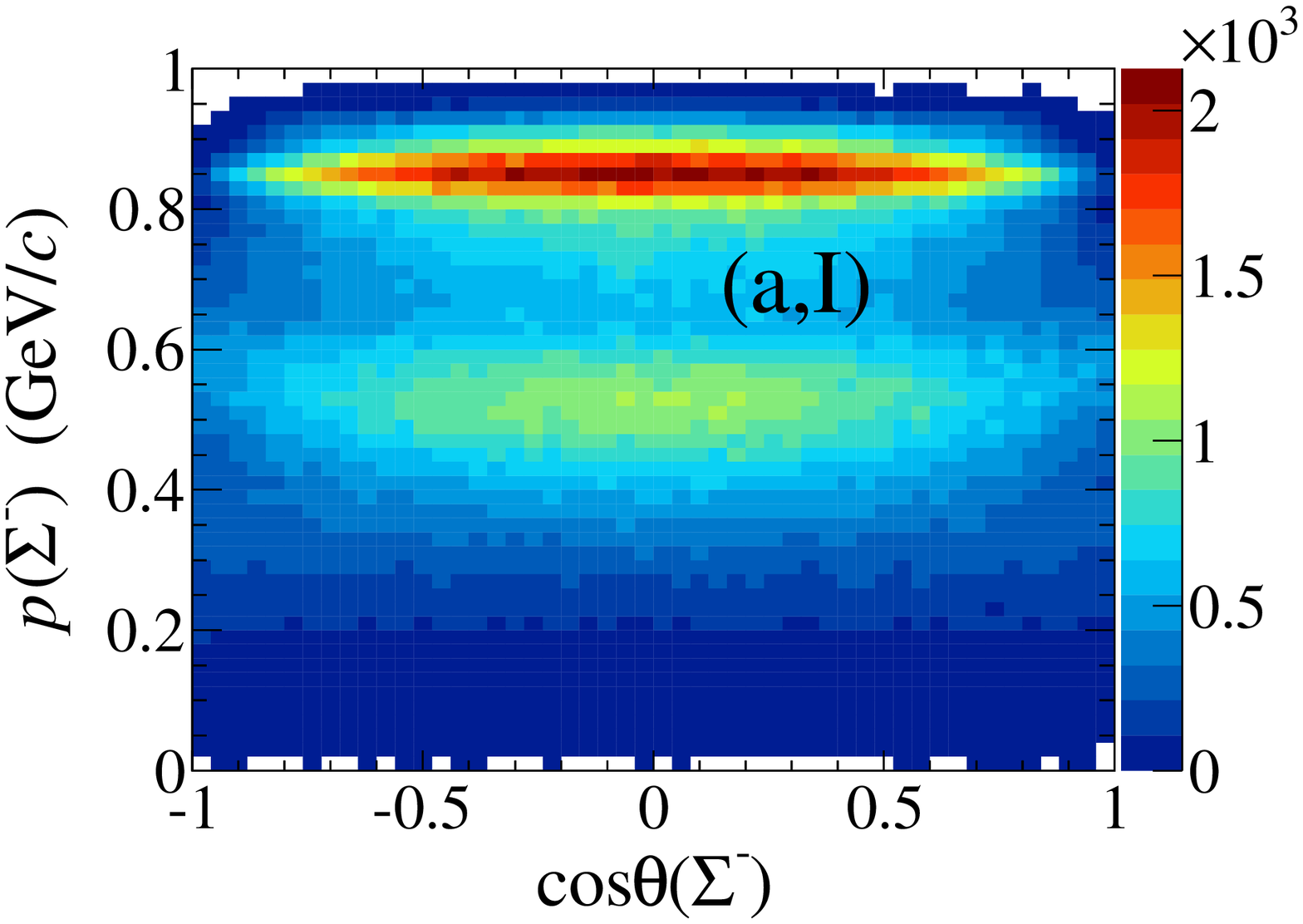}
        \includegraphics[width=2.0in]{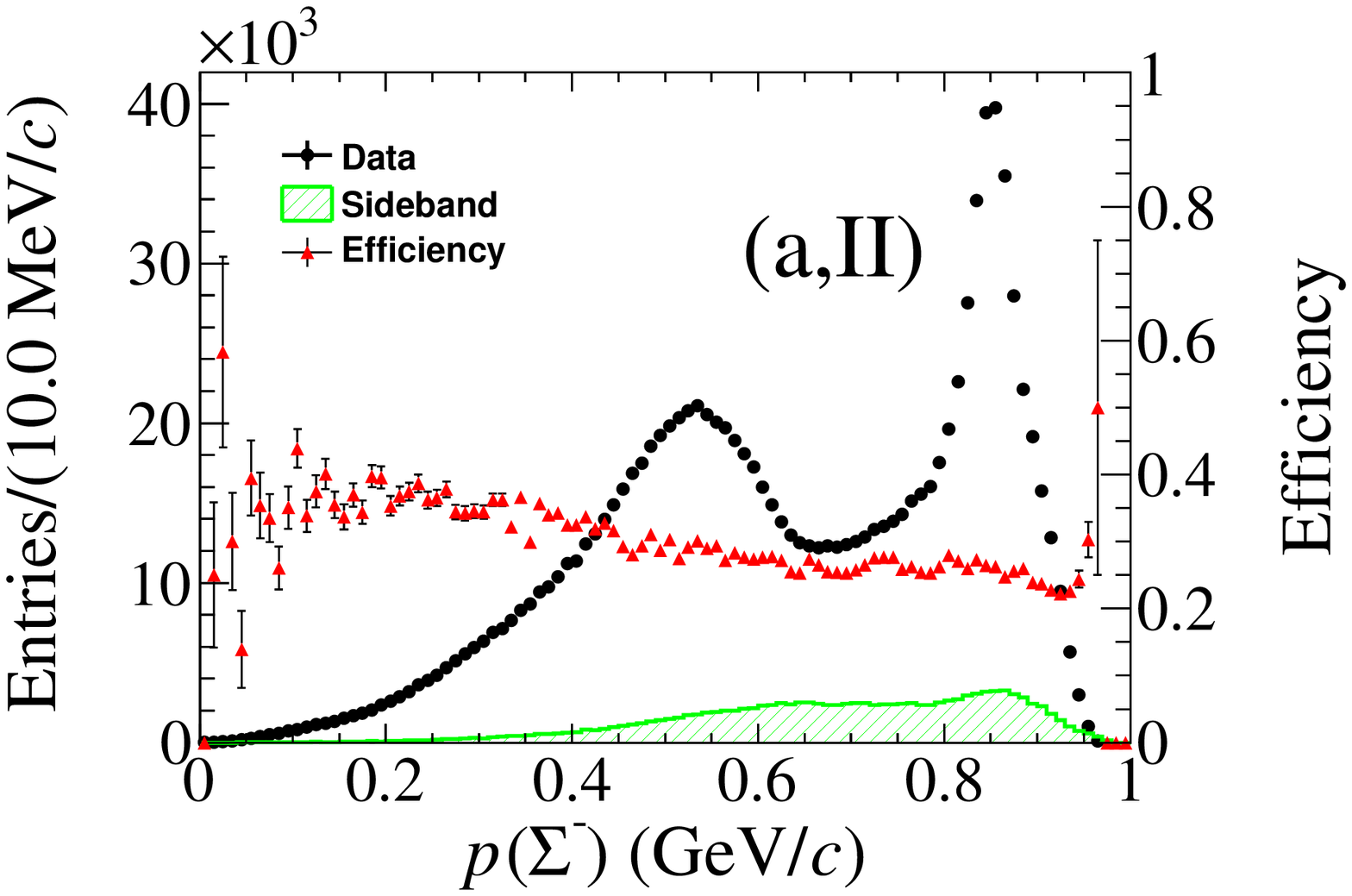}
        \includegraphics[width=2.0in]{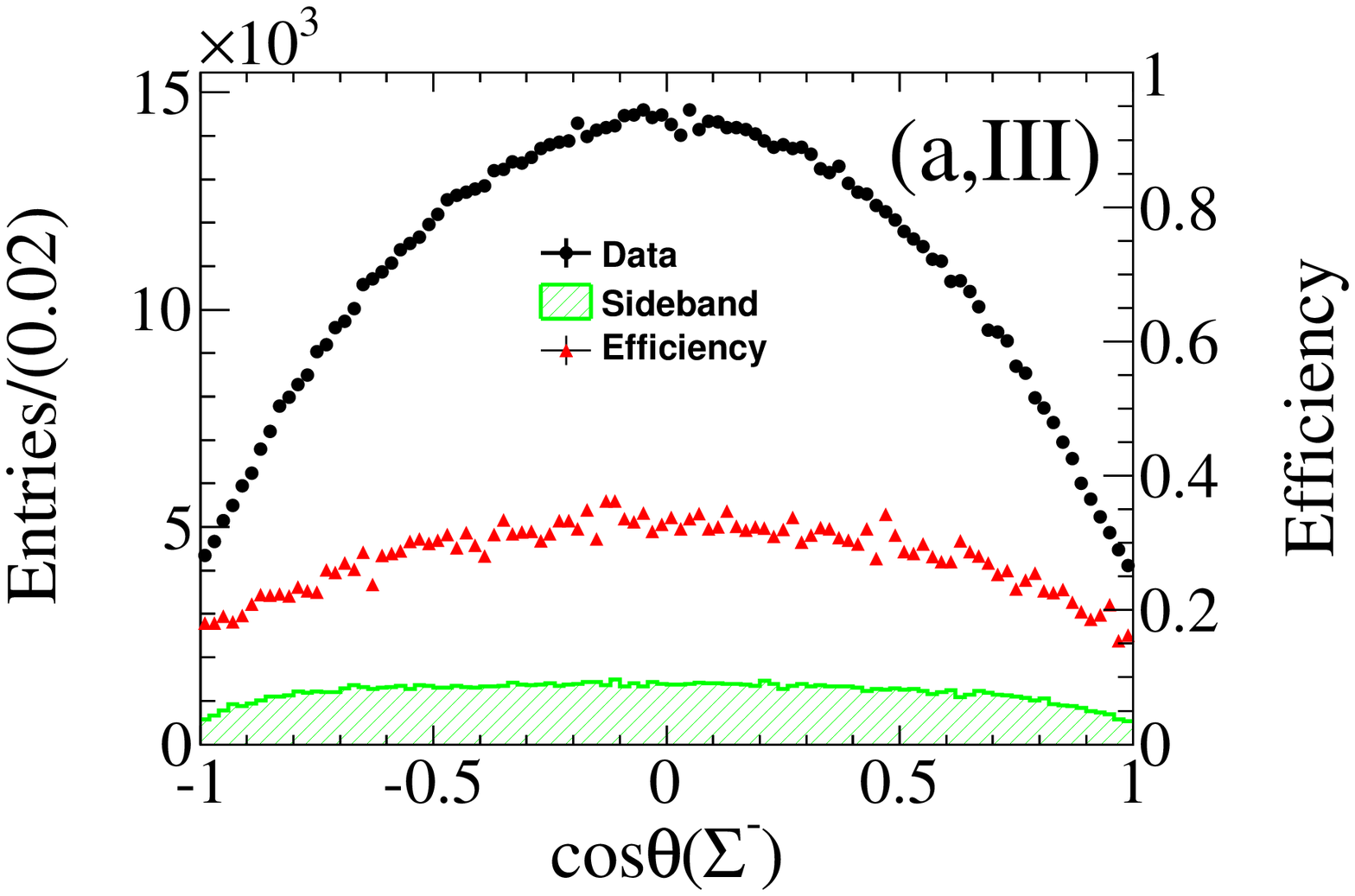}
        
        \includegraphics[width=2.0in]{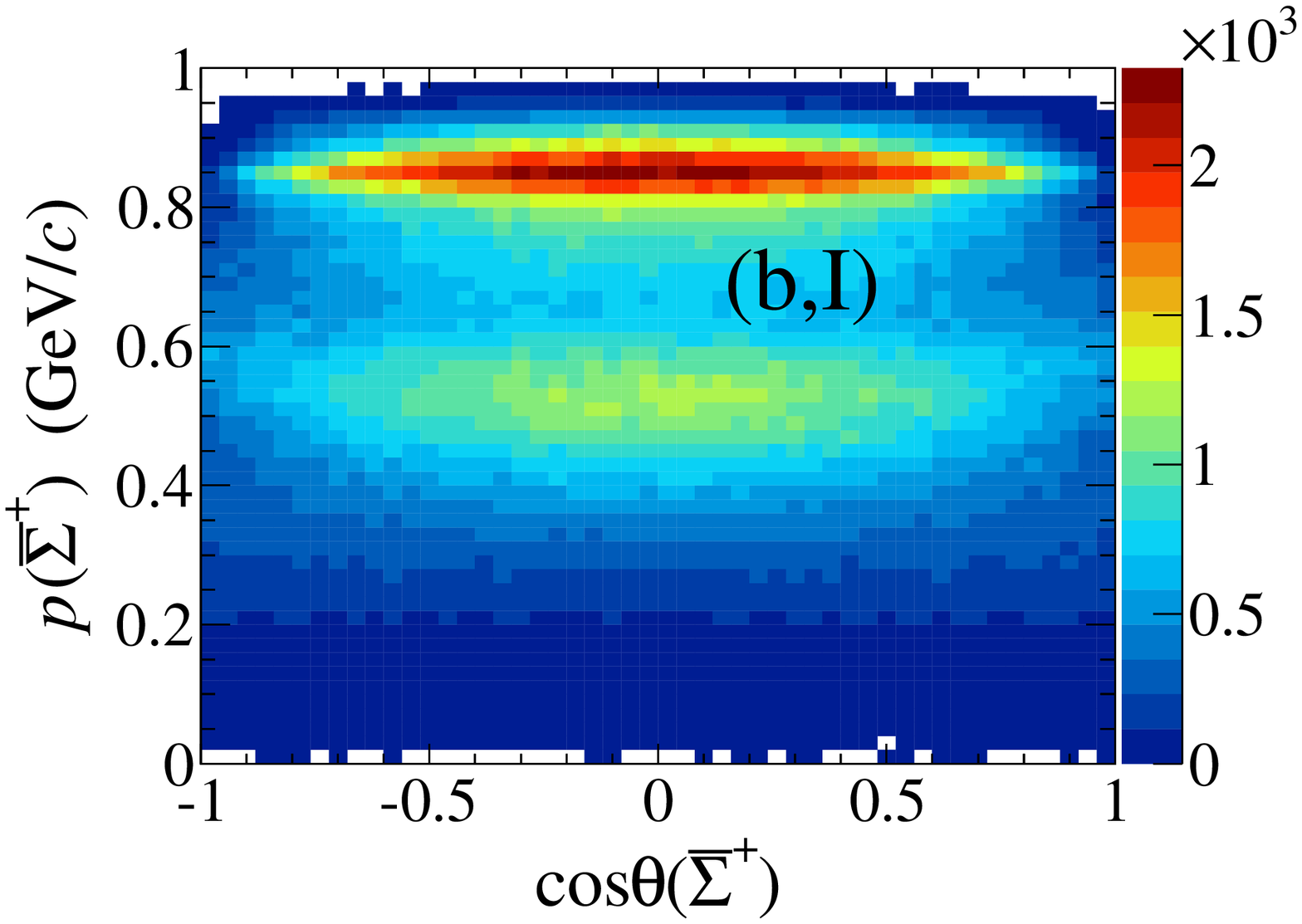}
        \includegraphics[width=2.0in]{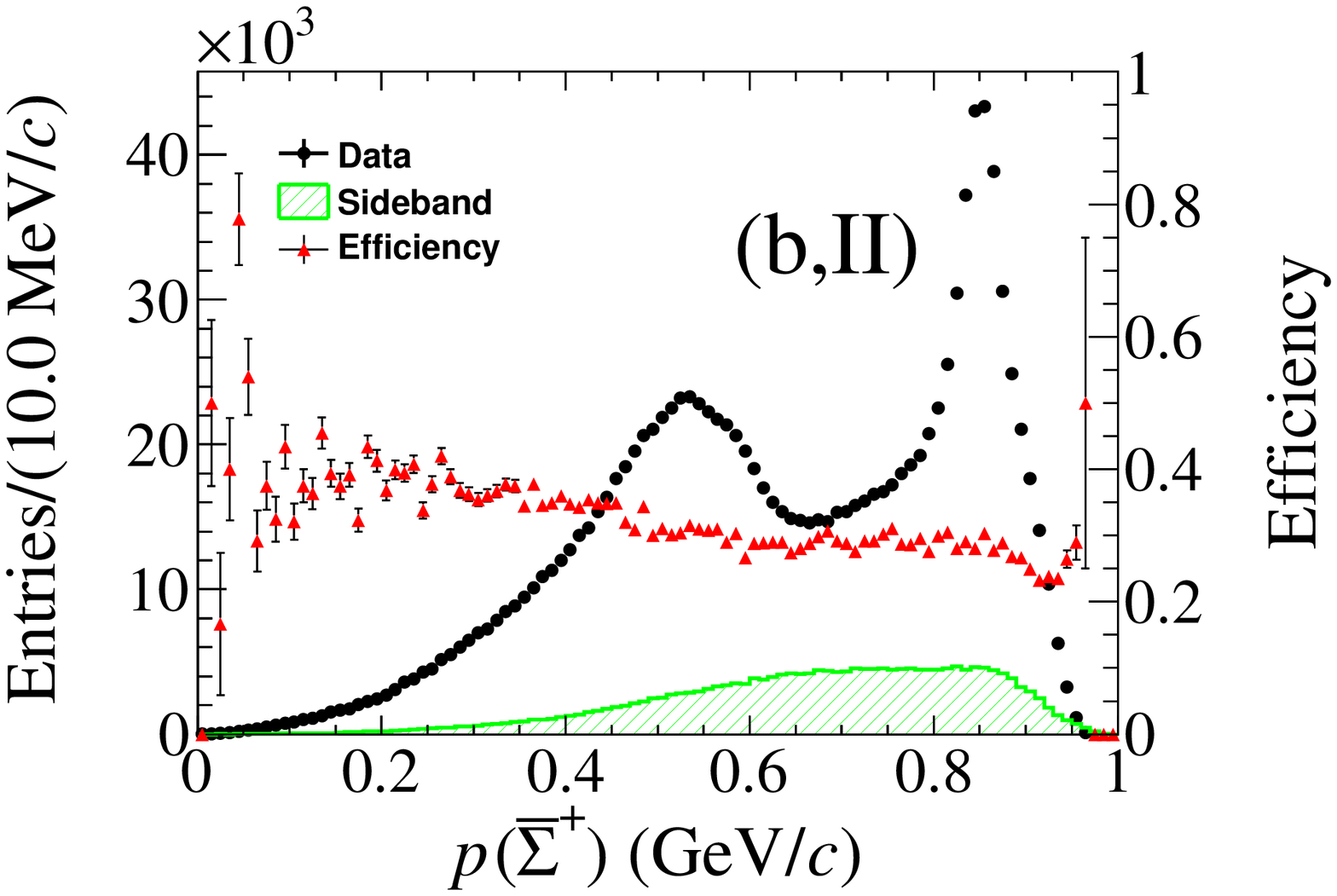}
        \includegraphics[width=2.0in]{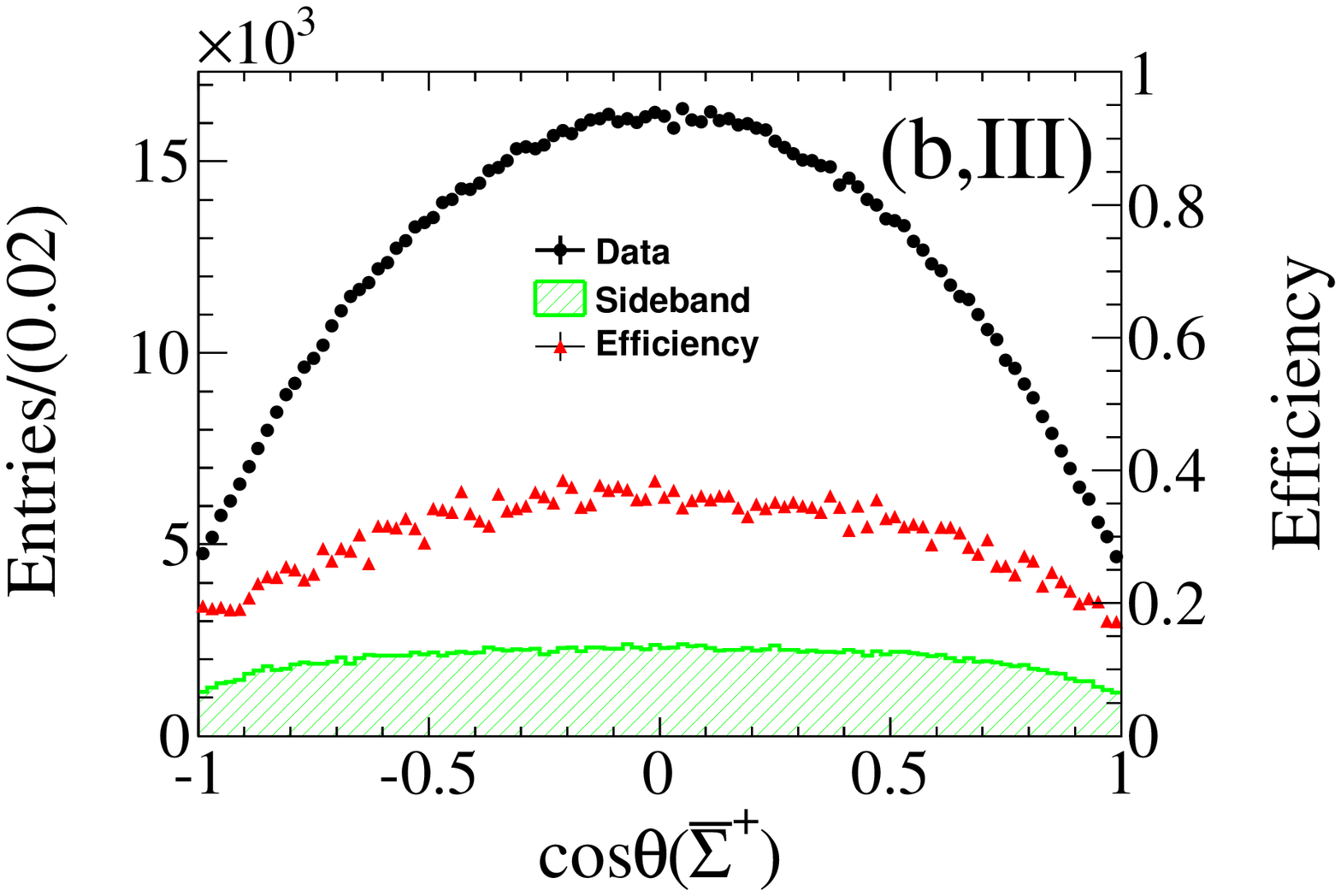}
        
        \includegraphics[width=2.0in]{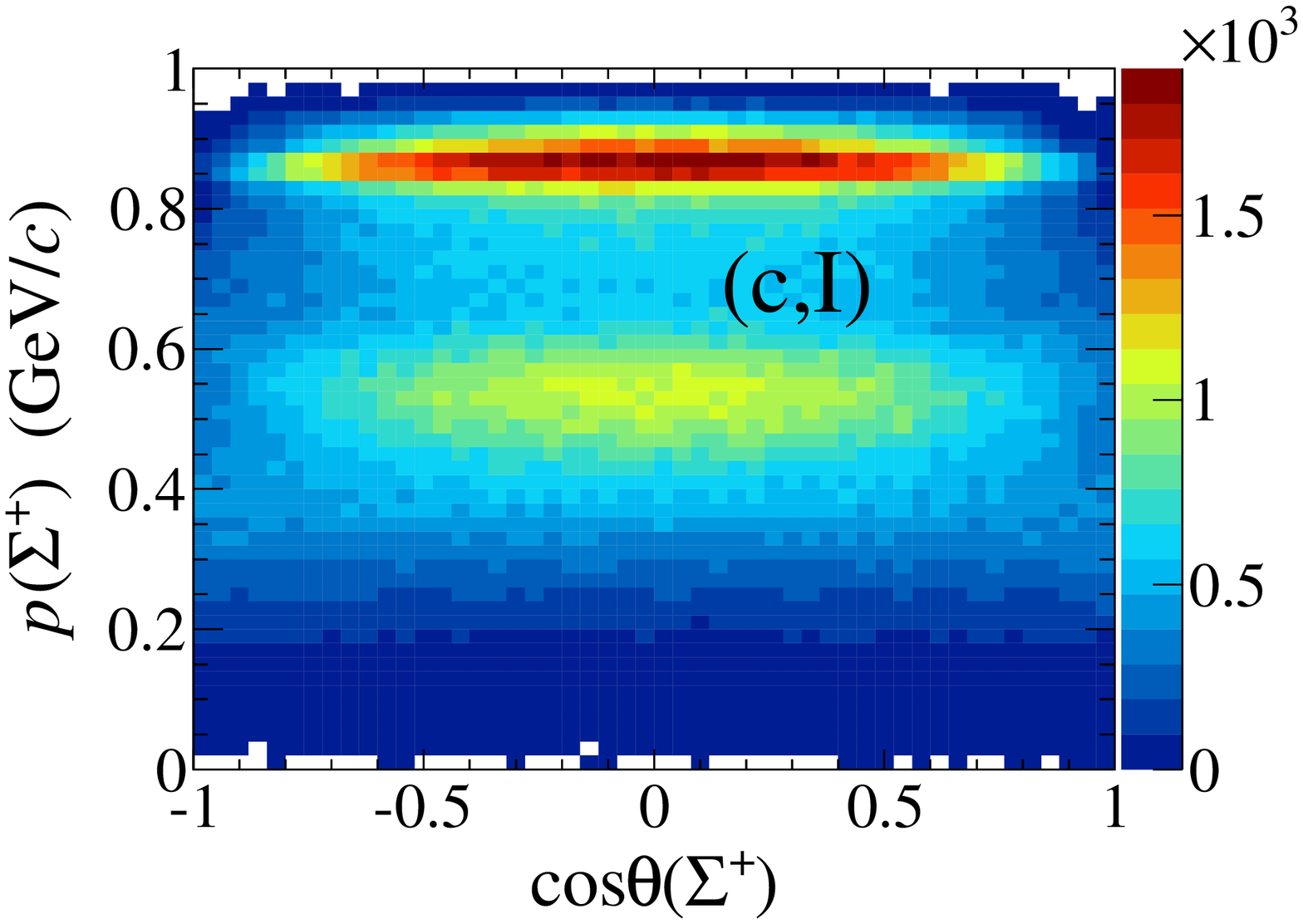}
        \includegraphics[width=2.0in]{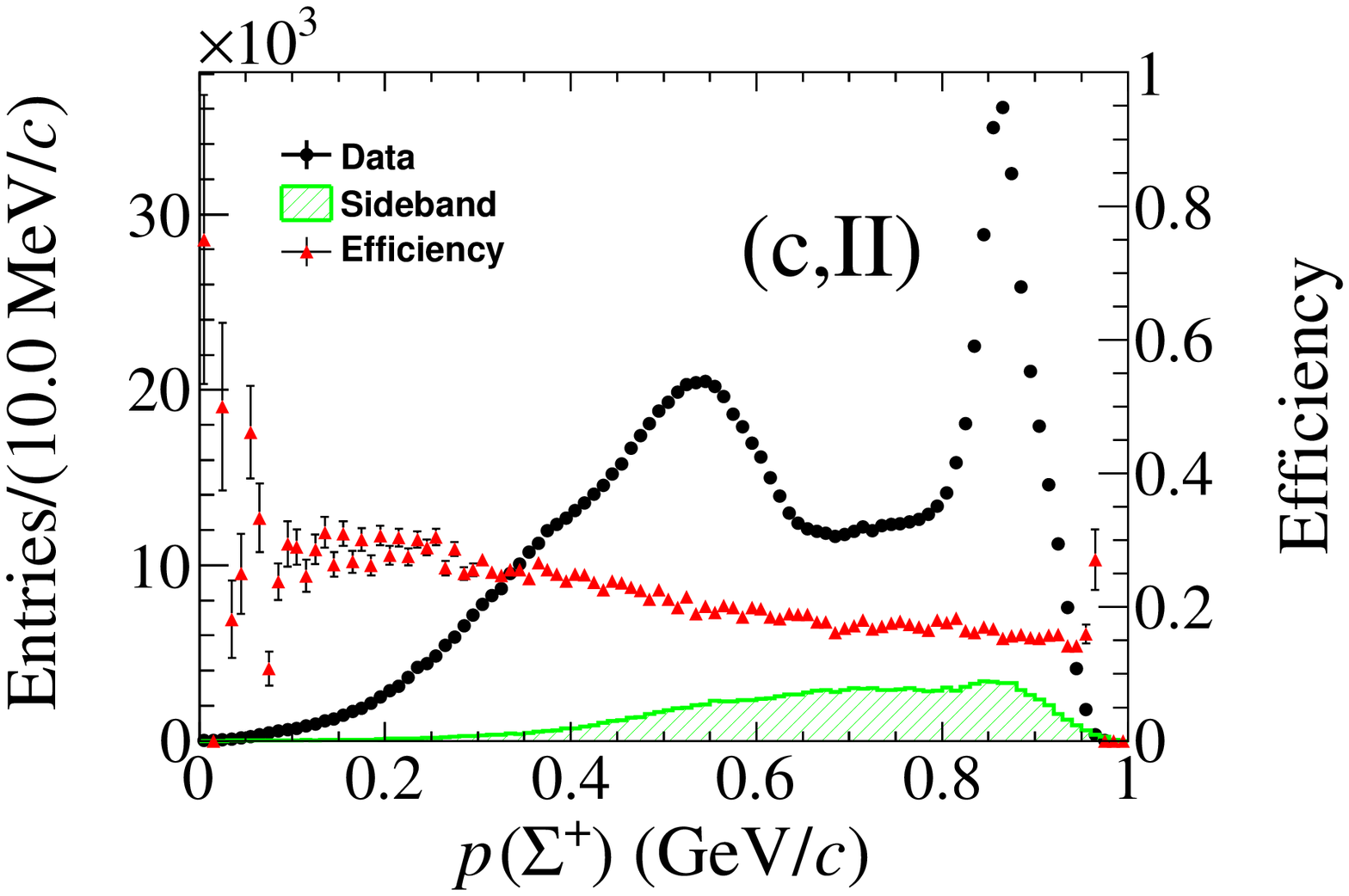}
        \includegraphics[width=2.0in]{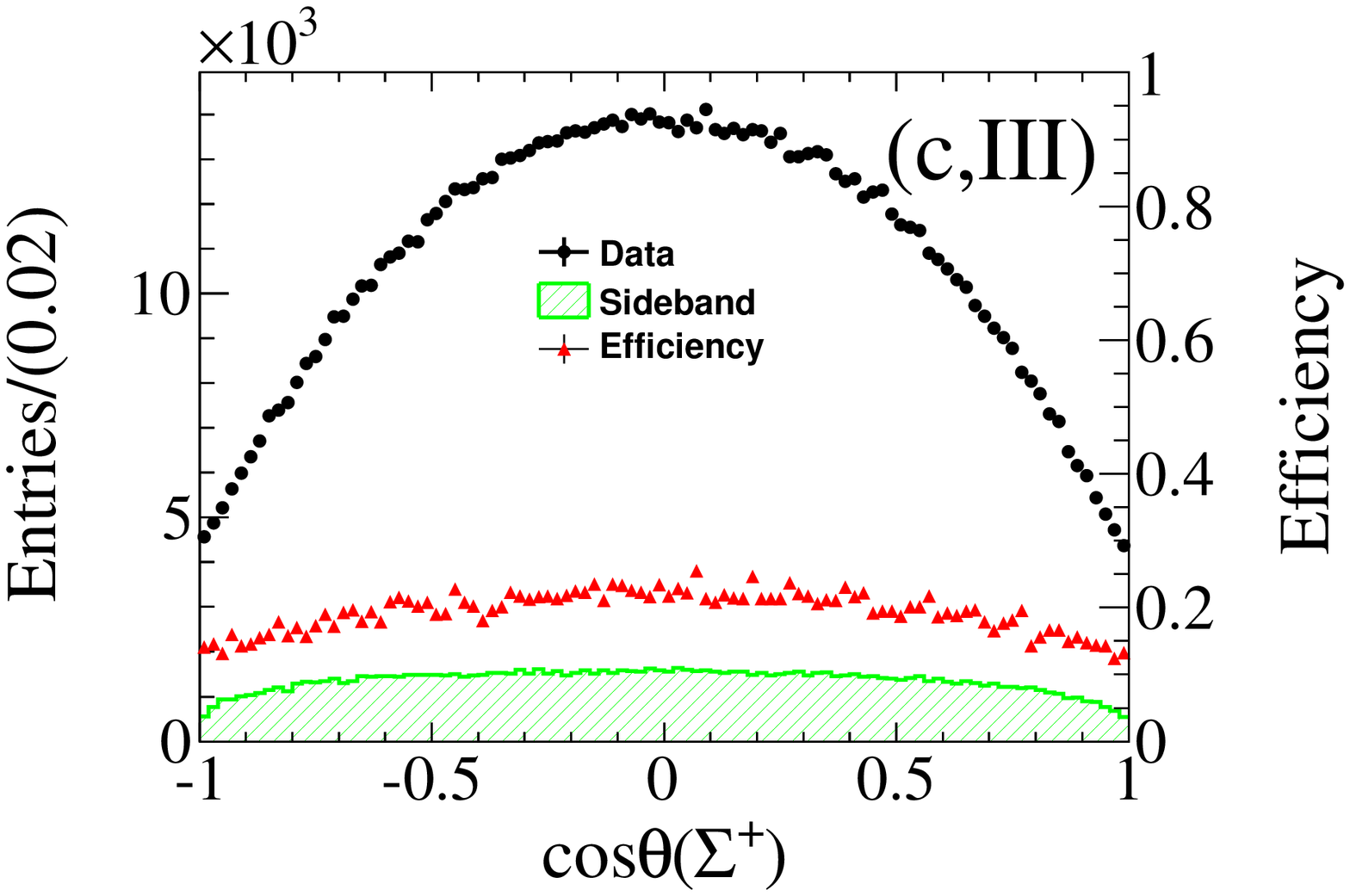}
        
        \includegraphics[width=2.0in]{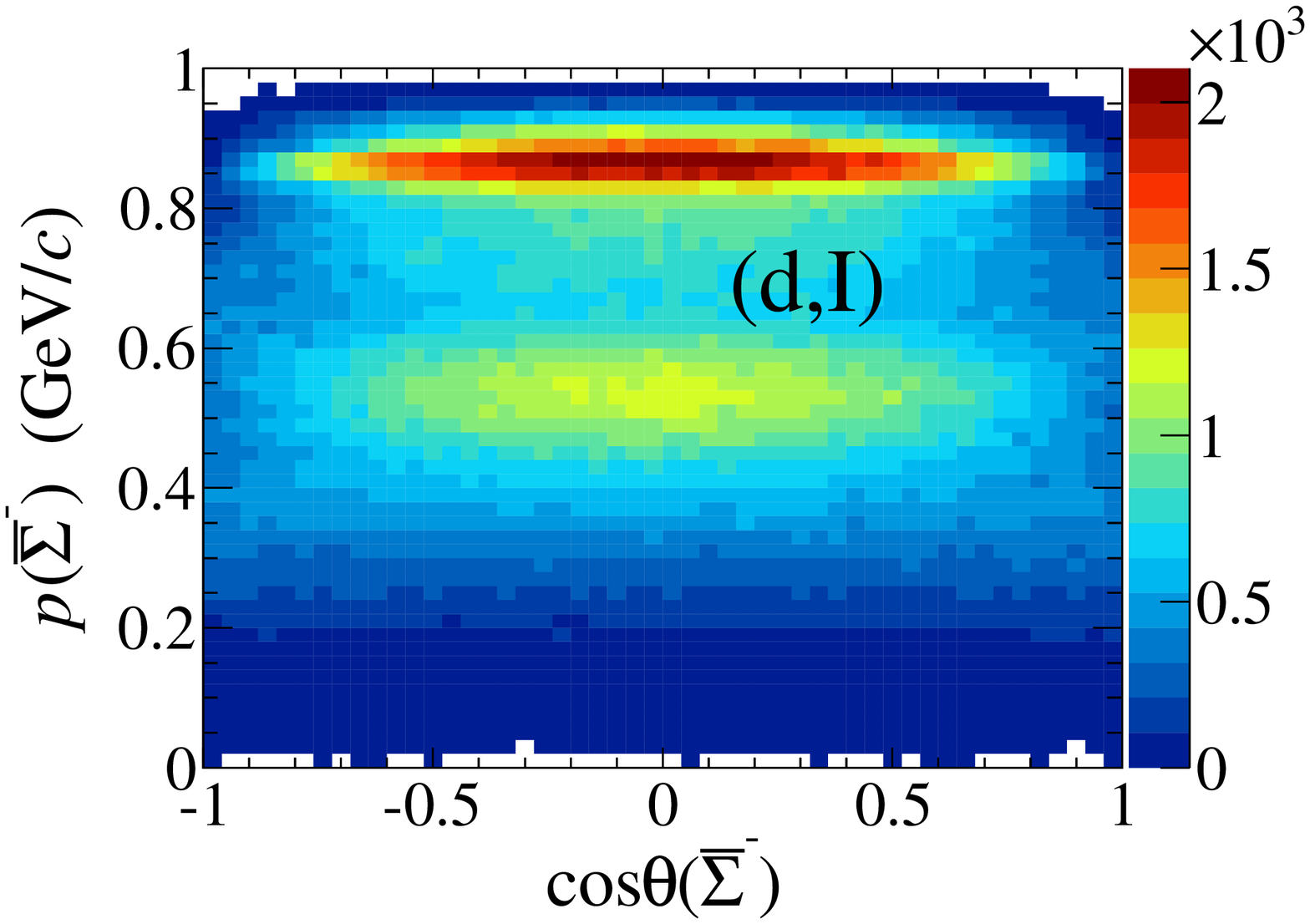}
        \includegraphics[width=2.0in]{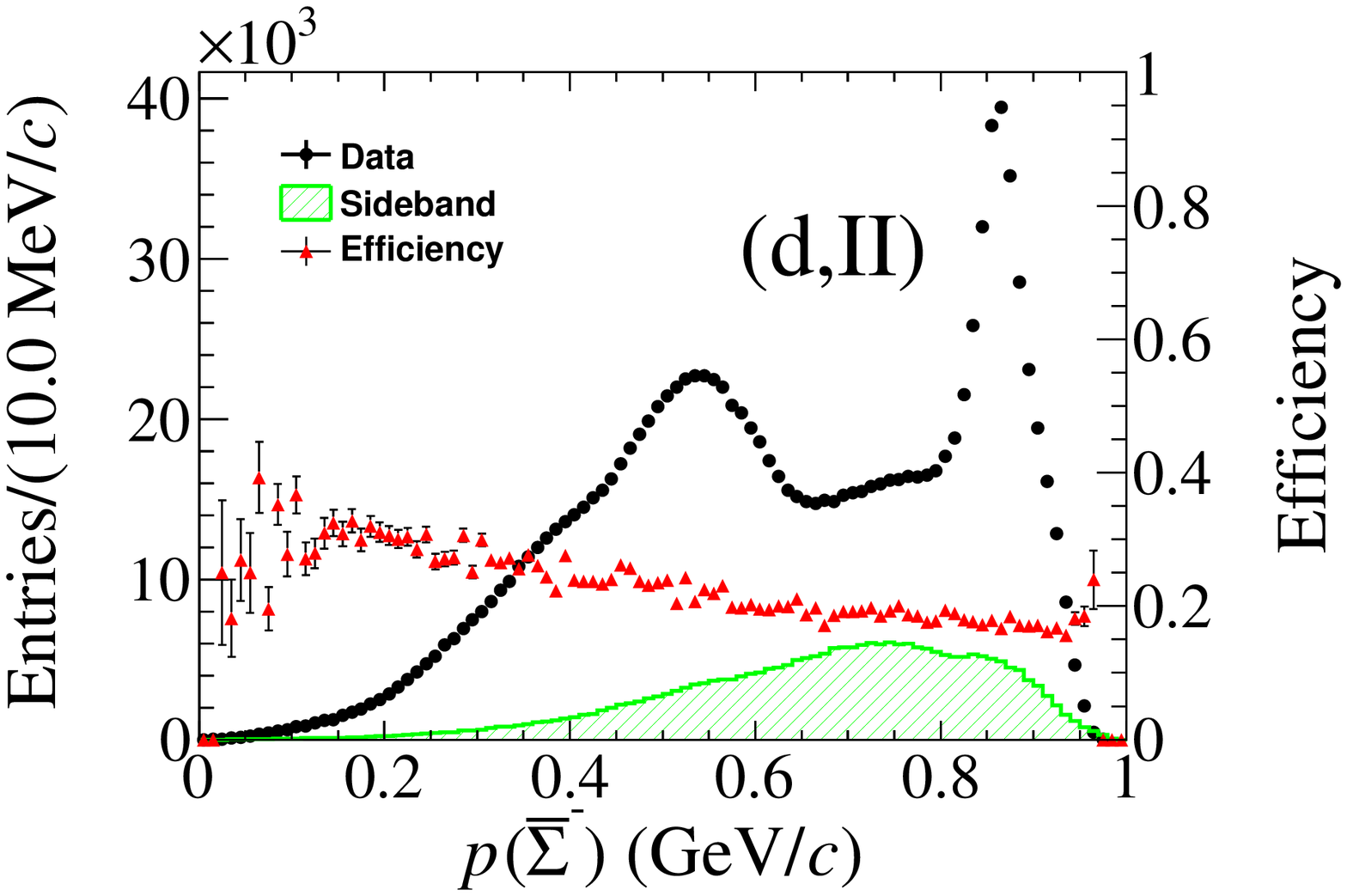}
        \includegraphics[width=2.0in]{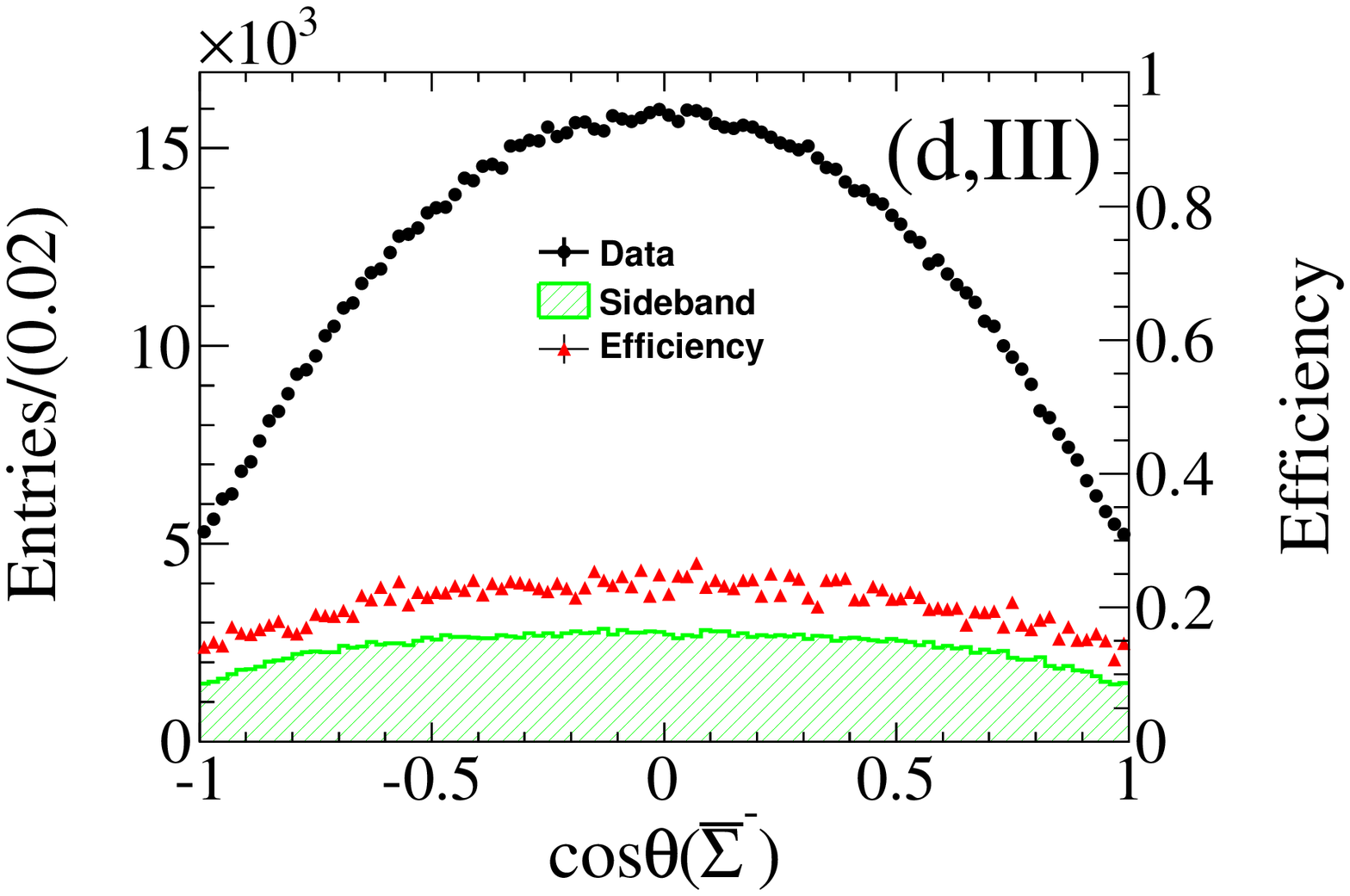}
        
        \caption{
            The distributions of $p$ versus $\cos\theta$ of $\Sigma$ sources from data in the $\Sigma$ signal region~(I), the momentum distribution of $\Sigma$ sources~(II), and the angular distribution~(III) from sources of $\Sigma^{-}$~(a), $\bar{\Sigma}^{+}$~(b), $\Sigma^{+}$~(c), and $\bar{\Sigma}^{-}$~(d).  The black dots with error bars are from data in the $\Sigma$ signal region. The green histograms are from data in the $\Sigma$ sideband regions. The red dots with error bars are the efficiencies of event reconstruction in each interval.
        }
        \label{fig:sigmasoure}
    \end{figure*}

    When $\bar{\Lambda}$ and $\pi^{\mp}$ are reconstructed successfully, $\Sigma$ signals can be found clearly in the recoiling side, which means we can control the $\Sigma$ baryon sources by selecting $\bar{\Lambda}\pi$ combinations. The events in $\Sigma^{\pm}$ signal regions can be used as $\Sigma$ sources to conduct fixed-target experiments, and those in the $\Sigma$ sideband regions can be used to estimate the backgrounds of the $\Sigma$ baryon sources. In our study of these four channels, the momentum and $\rm{cos\theta}$ distributions of the $\Sigma$ source particles are obtained in the laboratory frame, which is shown in Fig.~\ref{fig:sigmasoure}. Compared with fixed momentum particle sources from $\jpsi$ two-body decays~\cite{BESIII:2023clq}, the momentum range of the sources from three-body decays is wider, which means the cross section spectrum can be studied. The $\Sigma$  particles can be well identified with low background. By comparing the generated MC truth information and the values measured in the recoiling side, the momentum resolution of $\Sigma$ sources is better than 5~MeV/$c$ and the angular resolution is better than one degree. These $\Sigma$ particles can be used as sources in a future experiment to study their interaction with targets put close to the production vertex of the $\jpsi$ meson~\cite{CZYuan:2021,WMSong:2022,Dai:2022}. These sources will be very helpful for high-precision measurements of many $\Sigma$-nucleon reactions~\cite{Haidenbauer:2020, Rowley:2021, Miwa:2022, Song:2022}.

\acknowledgments

The BESIII Collaboration thanks the staff of BEPCII and the IHEP computing center for their strong support. This work is supported in part by National Key R\&D Program of China under Contracts Nos. 2020YFA0406300, 2020YFA0406400; National Natural Science Foundation of China (NSFC) under Contracts Nos. 11635010, 11735014, 11835012, 11935015, 11935016, 11935018, 11961141012, 12022510, 12025502, 12035009, 12035013, 12061131003, 12150004, 12192260, 12192261, 12192262, 12192263, 12192264, 12192265, 12221005, 12225509, 12235017; the Chinese Academy of Sciences (CAS) Large-Scale Scientific Facility Program; the CAS Center for Excellence in Particle Physics (CCEPP); CAS Key Research Program of Frontier Sciences under Contracts Nos. QYZDJ-SSW-SLH003, QYZDJ-SSW-SLH040; 100 Talents Program of CAS; The Institute of Nuclear and Particle Physics (INPAC) and Shanghai Key Laboratory for Particle Physics and Cosmology; ERC under Contract No. 758462; European Union's Horizon 2020 research and innovation programme under Marie Sklodowska-Curie grant agreement under Contract No. 894790; German Research Foundation DFG under Contracts Nos. 443159800, 455635585, Collaborative Research Center CRC 1044, FOR5327, GRK 2149; Istituto Nazionale di Fisica Nucleare, Italy; Ministry of Development of Turkey under Contract No. DPT2006K-120470; National Research Foundation of Korea under Contract No. NRF-2022R1A2C1092335; National Science and Technology fund of Mongolia; National Science Research and Innovation Fund (NSRF) via the Program Management Unit for Human Resources \& Institutional Development, Research and Innovation of Thailand under Contract No. B16F640076; Polish National Science Centre under Contract No. 2019/35/O/ST2/02907; The Swedish Research Council; U. S. Department of Energy under Contract No. DE-FG02-05ER41374.


\end{document}